\documentclass[aps,preprint,superscriptaddress]{revtex4}
\usepackage{amsfonts}
\usepackage{amsmath}
\usepackage{amssymb}
\usepackage{mathrsfs}
\usepackage{subfigure}
\usepackage{graphicx}
\usepackage{ulem}
\usepackage{epstopdf}
\usepackage{float}
\usepackage{color}
\usepackage{bm}
\usepackage{ulem}
\usepackage{makecell}
\usepackage{xcolor}
\usepackage{appendix}
\usepackage{booktabs}
\usepackage{hyperref}
\usepackage{dcolumn}
\usepackage{array}

\usepackage{changes}

\hypersetup{colorlinks=true, linkcolor=blue, filecolor=blue, urlcolor=blue, citecolor=blue}
\allowdisplaybreaks
\begin{document}
\title{ Black holes in $f(R,T)$ gravity coupled with Euler-Heisenberg electrodynamics}
\preprint{CTP-SCU/2025014}
\author{Yizhi Liang}
\email{liangyizhi@stu.scu.edu.cn}
\affiliation{Center for Theoretical Physics, College of Physics, Sichuan University, Chengdu, 610065, China}
\author{Jun Tao}
\email{taojun@scu.edu.cn (Corresponding Author)}
\affiliation{Center for Theoretical Physics, College of Physics, Sichuan University, Chengdu, 610065, China}
\author{Rui Yang}
\email{yangrui7@stu.scu.edu.cn}
\affiliation{Center for Theoretical Physics, College of Physics, Sichuan University, Chengdu, 610065, China}
\begin{abstract}
We investigate the scenario of black holes coupled with the Euler-Heisenberg nonlinear electromagnetic field in the framework of $f(R,T)$ gravity. The black hole solutions for electrically charged, magnetically charged and the dyonic case are separately analyzed, and we discuss the scalar curvature and the energy conditions of the black hole spacetime. In the magnetic charge solution, the $f(R,T)$ correction appears in the $1/r^6$ Euler-Heisenberg electromagnetic field correction term, while the electrically charged black hole solution exhibits an $r^2$ term in the metric function, corresponding to an effective cosmological constant $\Lambda_\text{eff} = \beta/(3\pi)$, inducing asymptotically anti-de Sitter or de Sitter spacetimes depending on the sign of the coupling parameter $\beta$. The dyonic solution is obtained through vacuum polarization quantum electrodynamics corrections, where electromagnetic duality is broken and the solution contains higher-order correction terms. The relationship between the event horizon and charge of dyonic extreme black holes is studied. Furthermore, we investigate the effective metric, photon trajectories, and innermost stable circular orbit under nonlinear electromagnetic effects, providing images of photon geodesic for varying electric and magnetic charge strengths and $f(R,T)$ coupling parameter. 
\end{abstract}
\maketitle
\section{Introduction}
\par Einstein's General Relativity (GR), as the cornerstone of modern gravitational physics, successfully predict many physical phenomena \cite{LIGOScientific:2016aoc,LIGOScientific:2017ync,Banados:1992wn}. In recent years, due to advanced observational precision, GR has encountered difficulties in explaining phenomena such as the accelerated expansion of the universe \cite{Lucchin:1984yf,Copeland:2006wr,Marsh:2015xka,Weinberg:2013agg}, dark energy and dark matter \cite{Planck:2018vyg,WMAP:2003elm,ParticleDataGroup:2014cgo,Copeland:2006wr,Clowe:2006eq}, necessitating the development of new theories to better describe the cosmology. The $f(R,T)$ gravity theory can be regarded as the maximal extension of the Hilbert-Einstein action and has been widely applied in cosmological studies, particularly addressing the issues of early universe inflation and late-time accelerated expansion. Compared to the $f(R)$ gravity (see Refs. \cite{DeFelice:2010aj,Nojiri:2017ncd,Nojiri:2010wj} for a review), the $f(R,T)$ gravity introduces a direct coupling between matter and geometry. Specific functional forms of $f(R,T)$ have been used to fit observational data, such as Type Ia supernova data  \cite{Siggia:2024pgr,SupernovaCosmologyProject:1997zqe,SupernovaCosmologyProject:1998vns} and cosmic microwave background(CMB) radiation \cite{Rudra:2020nxk}. Related studies include energy condition analyses \cite{Sharif:2012gz}, investigations of white dwarfs and exotic stars \cite{Carvalho:2017pgk,Bhattacharjee:2023olf,Deb:2018gzt,Pretel:2020oae}. In the field of quantum cosmology, $f(R,T)$ gravity has also been explored to study its quantum behavior under the Friedmann-Robertson-Walker metric \cite{Sebastiani:2016ras,Xu:2016rdf}, aiming to explain the early evolution of the universe \cite{Myrzakulov:2012qp,Li:2025rog}, etc.
\par Euler and Heisenberg proposed a nonlinear electrodynamics based on Dirac  theory \cite{Heisenberg:1936nmg}, it is an effective theory derived from quantum electrodynamics (QED) after one-loop quantization. The vacuum is treated as a medium with a dielectric constant, with its polarizability and magnetizability described as a cloud of virtual charges surrounding real currents and charges. Schwinger reformulated this one-loop effective Lagrangian within the QED framework \cite{Schwinger:1951nm}. The static black hole solutions in the Einstein-Euler-Heisenberg (EEH) framework with electric, magnetic, and dyonic charges are investigated \cite{Magos:2023nnb,Ruffini:2013hia,Meng:2021cgb}, nonlinear effects introduce QED corrections to the black hole horizon, entropy, total energy, and maximum extractable energy \cite{Amaro:2020xro,Amaro:2022yew,Amaro:2022del,Breton:2023bwf,Amaro:2023ull,Cotton:2021tfl}.  Experimentally, when the electromagnetic field strength is high, the EH theory provides a more accurate classical approximation of QED compared to Maxwell theory \cite{Brodin:2001zz,Boillat:1970gw,Stehle:1966wii}. The EH action breaks electromagnetic duality at higher orders of the electromagnetic field, leading to significant distinctions in black hole solutions with electric or magnetic charges \cite{Gibbons:1995cv}.  From a modern perspective, string theory and D-brane physics, in the low-energy limit, give rise to Abelian and non-Abelian Born-Infeld (BI) lagrangians \cite{Fradkin:1985qd,Abouelsaood:1986gd,Tseytlin:1997csa}. Subsequent work has extensively studied the asymptotically flat, static, spherically symmetric black hole solutions of the Einstein-BI theory \cite{GarciaD:1984xrg,Demianski:1986wx}. As a low-energy limit of the BI theory, under appropriate parameters, the EH action effectively approximates the supersymmetric system of minimally coupled particles with spins -1/2 and 0 \cite{Bern:1993tz}. Ref. \cite{Yajima:2000kw} constrains the nonlinear parameter \( a > 0 \), relating it to the inverse string tension \( \alpha' \), and studies the effective EH Lagrangian as the low-energy limit of BI black hole solutions \cite{Allahyari:2019jqz,Vagnozzi:2022moj,Li:2021ygi,Meng:2021cgb,Magos:2020ykt,Blazquez-Salcedo:2020caw,Karakasis:2022xzm,Maceda:2018zim,Gutierrez-Cano:2024oon,Rehman:2023hro,Breton:2021mju}. 

 \par The Reissner-Nordstr\"{o}m(RN) metric is an exact solution to the Einstein-Maxwell equations such that the algebraic structures of the gravitational and electromagnetic fields are fully aligned \cite{Podolsky:2025tle}. Furthermore, electric charge has been postulated to facilitate the stabilization of wormhole throats, potentially rendering them traversable \cite{Konoplya:2021hsm, Blazquez-Salcedo:2020czn}. In astrophysical contexts, black holes are typically immersed in magnetic fields, a configuration that underpins the high energy activities observed in galactic nuclei and quasars. While the strong-field regimes near black holes may induce divergences within classical electrodynamics, Euler-Heisenberg electrodynamics provides a quantum electrodynamics (QED) approximation \cite{Nozari:2025zlt}. Investigating $f(R, T)$ gravity coupled with EH electromagnetic fields is valuable. Black holes with three different couplings—magnetic charge, electric charge, and dyonic charge—are studied, but using different viewpoints. For the magnetic and electric charge solutions, we adopt the Born-Infeld low-energy limit, reflecting their physical correspondence to the effective field theory of D-branes. The parameter $a$ is identified with the inverse string tension, thereby permitting a broader parameter space than the constrained QED interpretation \cite{Yajima:2000kw}; the QED one-loop corrected effective action is used for the dyonic solution, as the BI approach fails to adequately account for the electromagnetic duality breaking at higher-order terms in EH, with QED emphasizing quantum corrections that are suitable for describing dyonic systems where electromagnetic interactions dominate. Since the dyonic solution closely resembles the real environment around black holes, these quantum corrections are probably to produce observable effects.
\par The Event Horizon Telescope (EHT) collaboration successfully obtained the high-resolution images of the supermassive black hole Sgr A* \cite{SgrA_1_2022,SgrA_2_2022,SgrA_3_2022,SgrA_4_2022,SgrA_5_2022,SgrA_6_2022}. The observation indicates that this black hole exhibits a ring-like structure analogous to the M87* black hole released in 2019 – featuring a central dark region encircled by a bright emission ring \cite{M87_1_2019,M87_2_2019,M87_3_2019,M87_4_2019,M87_5_2019,M87_6_2019}.
General relativity predicts significant deflection of photons traversing strong gravitational fields: some photons are captured by the black hole, forming the shadow region, while some escapes and form photon rings around the black hole \cite{Shoom:2017ril}.
Theoretical investigations of black hole shadows begins from Synge's work on the Schwarzschild spacetime \cite{Synge:1966okc}. Bardeen et al. subsequently extended this research to the Kerr black hole scenario \cite{Bardeen:1972fi}. Luminet pioneered the numerical simulation of shadows for rotating black holes with accretion disks \cite{Luminet:1979nyg}. Contemporary theoretical models, incorporating radiated emissions from accreting matter, provide robust explanations for the EHT observations of both M87* and SgrA* \cite{Dokuchaev:2019pcx,Dokuchaev:2020wqk,Fang:2024hbw}.
Recent research has expanded into richer theoretical frameworks, including: Born-Infeld spacetimes \cite{Wen:2022hkv,He:2022opa}, perfect fluid dark matter models \cite{Yang:2024ulu}, Gauss-Bonnet AdS spacetimes \cite{Han:2018ooi},  hairy Schwarzschild black holes \cite{Meng:2023htc}, and black hole systems exhibiting multiple photon spheres \cite{Guo:2021enm,Chen:2022qrw,Guo:2022ghl,Guo:2022muy,Weng:2025fib,Nojiri:2017kex,Volonteri:2010wz}. Given that EHT imaging of M87* and Sgr A* probes photon behavior in extreme gravity and nonlinear electrodynamics, it is therefore imperative to investigate the behavior of photons around black holes in the $f(R,T)$-EH framework. This study examines how these couplings modify the effective metric, photon geodesics, and the innermost stable circular orbit (ISCO), potentially yielding observable shadow sizes and ring structures that deviate from GR predictions.
\par The structure of this paper is as follows: In Section \ref{s2}, we review $f(R,T)$ gravity theory, present the modified $f(R,T)$-EH action, and derive the field equations under minimum coupling $f(R,T)$. In Section \ref{s3}, we provide the black hole solution for the magnetic monopole EH field and discuss the nature of singularities. Section \ref{s7} electrically charged black holes is considered. Both Section \ref{s3} and Section \ref{s7} employ the Euler-Heisenberg interpretation of BI electrodynamics in low-energy limit. In Section \ref{s4}, we use the EH explanation based on vacuum charge polarization to analyze the dyonic black hole solution. In Section \ref{s5}, energy conditions are qualitatively analyzed. In Section \ref{s6}, we thoroughly investigate the effective metric, photon trajectories, and innermost stable circular orbit (ISCO) under nonlinear electromagnetic effects, providing images of photon geodesic integrals for varying electric and magnetic charge strengths and $f(R,T)$ coupling parameters. Finally, Section \ref{s8} summarizes the paper. For simplicity, we set the natural units $c=G_n=\hbar=1$ throughout the paper, and black hole mass is fixed at $M=1$.
\section{Field equations of $f(R, T)$ gravity coupled with EH electrodynamics\label{s2}}
\par According to Ref. \cite{Harko:2011kv} and combining the EEH action \cite{Yajima:2000kw}, we assume the EH nonlinear electromagnetic field as the matter section in the action, and this gravitational system can be described as
\begin{equation}
	S = \frac{1}{16\pi} \int f(R, T) \sqrt{-g} \, d^4x + \int \mathcal{L}(F) \sqrt{-g} \, d^4x \,,
\end{equation}
where we have $g=\det(g_{\mu\nu})$, $f(R, T)$ is an arbitrary function of the scalar curvature $R$ and of the energy-momentum tensor trace $T$, $\mathcal{L}(F)$ is the Lagrangian density of EH field, which has the form \cite{Yajima:2000kw,Ma:2024oqe}
\begin{equation}
	\mathcal{L}(F)=\frac{1}{4\pi}(-F+a F^2+b G^2),
\end{equation}
the invariants are defined as $F\equiv \frac{1}{4}F^{\mu\nu}F_{\mu\nu}$, $G\equiv\frac{1}{4}F_{\mu\nu}*F^{\mu\nu}$, where the $*F^{\mu\nu}$ is the dual form of the electromagnetic strength tensor $F_{\mu\nu}$.
 The parameters $a, b$ are the coupling constants due to the weak field approximation and we have $a=he^4/(360\pi^2m_e^4)\approx0.001768, b = 7he^4/(1440\pi^2m_e^4)\approx0.003095$, $h$ is the Plank constant, $e, m$ are the charge and the mass of electron, respectively \cite{Heisenberg:1936nmg}. For BI low energy limit, the EH allows parameters $a,b$ to be interpreted as the inverse string tension $\alpha^\prime$, which means they are free parameters. Varying this action with respect to $ g^{\mu\nu}$ and gauge field $A_\mu$, we can obtain the equation of motion
\begin{equation}
	\begin{split}
		\delta S = & \frac{1}{16\pi} \int \left[ f_R(R,T) R_{\mu\nu} \delta g^{\mu\nu} + f_R(R,T) g_{\mu\nu} \Box \delta g^{\mu\nu} + \right. \\
		& \left. - f_R(R,T) \nabla_\mu \nabla_\nu \delta g^{\mu\nu} + f_T(R,T) \frac{\delta (g^{\alpha\beta} T_{\alpha\beta})}{\delta g^{\mu\nu}} \delta g^{\mu\nu} + \right. \\
		& \left. - \frac{1}{2} g_{\mu\nu} f(R,T) \delta g^{\mu\nu} + \frac{16\pi}{\sqrt{-g}} \frac{\delta (\sqrt{-g} \mathcal{L}_D)}{\delta g^{\mu\nu}} \right] \sqrt{-g} d^4 x,
	\end{split}
\end{equation}
\begin{equation}
	\frac{1}{\sqrt{-g}} \partial_\mu \left[ \sqrt{-g} \left( 2 f_T(R,T) \mathcal{L}_{FF} F -  \mathcal{L}_F \right) F^{\mu\nu} \right] = 0,
\end{equation}
where $\mathcal{L}_F=\partial\mathcal{L}/\partial F, \mathcal{L}_{FF}=\partial^2\mathcal{L}/\partial F^2$. Yields the modified Einstein field equation reads
\begin{equation}
	G_{\mu\nu}\equiv f_R R_{\mu\nu} + \left( g_{\mu\nu} \Box - \nabla_{\mu} \nabla_{\nu} \right) f_R - \frac{1}{2} f g_{\mu\nu} = T_{\mu\nu} - f_T (T_{\mu\nu} + \Theta_{\mu\nu}),
\end{equation}
here we have $f_R = \partial f/\partial R$, $f_T = \partial f/\partial T$ and$\Box= \nabla_\rho \nabla^\rho$. The energy-momentum tensor $T_{\mu\nu}$ and $\Theta_{\mu\nu}$ reads
\begin{equation}
	T_{\mu\nu} \equiv -\frac{2}{\sqrt{-g}} \frac{\delta (\sqrt{-g} \mathcal{L}(F)}{\delta g^{\mu\nu}},
	\label{T}
\end{equation}

\begin{equation}
	\Theta_{\mu\nu} \equiv g^{\alpha\beta} \frac{\delta T_{\alpha\beta}}{\delta g^{\mu\nu}}= -2T_{\mu\nu} + g_{\mu\nu} \mathcal{L}(F) - 2g^{\alpha\beta} \frac{\partial^2 \mathcal{L}(F)}{\partial g^{\mu\nu} g^{\alpha\beta}},
	\label{theta}
\end{equation}
using the result mentioned in Ref. \cite{Weinberg:1972kfs}
\begin{equation}
	\delta \ln|\det -g|=-Tr(g^{\mu\nu}\delta g_{\mu\nu}),
\end{equation}
then the Eq. (\ref{T}) and Eq. (\ref{theta}) can be expressed as
\begin{equation}
	\begin{split}
		T_{\mu\nu}&=g_{\mu\nu} \mathcal{L}(E)-\mathcal{L}_F F_{\mu\rho}F_\nu^\rho,\\
		\Theta_{\mu\nu}&=-g_{\mu\nu}\mathcal{L}(E)+\mathcal{L}_F F_{\mu\rho}F_\nu^\rho-2\mathcal{L}_{FF}FF_{\mu\rho}F_\nu^\rho,
	\end{split}
	\label{9}
\end{equation}
taking the trace of the Einstein equation, we can derive
\begin{equation}
	\Box f_R =\frac{1}{3} \left( T - f_T (T + \Theta) + 2f - f_R R \right).
\end{equation}
We choose the linear coupling form \cite{Harko:2011kv} $f(R,T)=R+2\beta T$, and the field equation can be expressed as
\begin{equation}
	R_{\mu\nu}+\frac{1}{2}Rg_{\mu\nu}=8\pi T_{\mu\nu}-2\beta(T_{\mu\nu}+\Theta_{\mu\nu})+\beta T g_{\mu\nu}.
	\label{11}
\end{equation}
Noticed that when $\beta=0$, it reduces to Einstein gravity. The product of the energy-momentum tensor trace $T$ and the metric $g_{\mu\nu}$ indicates that matter evolution affects spacetime geometry, similar to dark energy models, explaining cosmic acceleration. In the early universe, the evolution of $T$ under high-energy scales significantly alters the inflation \cite{Myrzakulov:2012qp}. This paper focuses on the spacetime structure coupled with the EH nonlinear electromagnetic field, discussed in detail in the following section.
\section{Magnetically charged black holes\label{s3}}
\par The idealized cases with purely magnetic and purely electric fields are considered. In these models, the EH theory is interpreted as the low-energy limit of the BI electrodynamics, and our results are compared with the electrically charged Euler–Heisenberg (EEH) solutions \cite{Yajima:2000kw}. Under this framework, the magnetic monopole model introduces magnetic charges similar to Dirac magnetic monopoles \cite{Nepomechie:1984wu}, describing static, spherically symmetric black hole solutions.  We have $Q_e\equiv0$, and the gauge potential is $A_\mu=(0,0,0,Q_m \cos\theta)$. The nonzero components of $F_{\mu\nu}$ and the invariants $F, G$ are
\begin{equation}
	F_{\theta\phi}=-F_{\phi\theta}=Q_m \sin\theta, \quad F=\frac{Q_m^2}{2r^4},\quad G=0.
\end{equation}
One can set the ansatz of the static and spherical symmetric black hole solution
\begin{equation}
	ds^2=-A(r)dt^2+B(r)dr^2+r^2d\Omega^2,
\end{equation}
where $d\Omega^2=r^2 d\theta^2+r^2\sin^2\theta d\phi^2$. The Ricci scalar can be expressed as
\begin{equation}
	R =g^{\mu\nu}R_{\mu\nu}= \frac{r^2 B A'^2 + r A \left(r A' B' - 2 B (r A'' + 2 A')\right) + 4 A^2 \left(r B' + B^2 - B\right)}{2 r^2 A^2 B^2},
\end{equation}
due to the spherical symmetry we have $A(r)=B(r)^{-1}$, it reduce to 
\begin{equation}
	R = -\left( A''(r) + \frac{4A'(r)}{r} + \frac{2A(r)}{r^2} - \frac{2}{r^2} \right),
\end{equation}
the prime denotes the derivative to $r$.
By substituting Eq. (\ref{9}) into Eq. (\ref{11}), we obtain the component field equation
\begin{align}
			-\frac{A(r) \left(r A'(r)+A(r)-1\right)}{r^2} &=-(8\pi+4\beta)A(r)\mathcal{L}(F)+4\beta A(r)\mathcal{L}_F F,\label{feq1}\\
		\frac{r A'(r)+A(r)-1}{r^2 A (r)} &=(8\pi+4\beta)\frac{1}{A(r)}\mathcal{L}(F)-4\beta \frac{1}{A(r)}\mathcal{L}_F F,\label{feq2} \\
		\frac{1}{2} r \left(r A''(r)+2 A'(r)\right) &=(8\pi+4\beta)r^2\mathcal{L}(F)+F_{\mu\rho}F_\nu^\rho(-8\pi\mathcal{L}_F+4\beta\mathcal{L}_{FF}F)-4\beta r^2\mathcal{L}_F F,	\label{feq3}
\end{align}
and notice the Eq. (\ref{feq1}) and Eq. (\ref{feq2}) are equivalent. The metric function can be expressed as
\begin{equation}
	A_m(r)=1-\frac{2M}{r}+\frac{Q_m^2}{r^2}-\frac{a Q_m^4}{10  r^6}+\frac{\beta a Q_m^4}{20 \pi  r^6}.
	\label{am}
\end{equation}
When $\beta\rightarrow 0$ we recover the standard EH solutions, the metric function for a magnetically charged black hole includes a correction term of $-a Q_m^4 / (10 r^6)$, which arises from the nonlinear electromagnetic effects.This term screens the magnetic charge at small radii, leading to modifications in the near-horizon geometry. When $\beta$ does not vanish, the $f(R,T)$ coupling effectively rescales the EH parameter by a factor of $(1-\beta/(2\pi))$. For $\beta<0$, the $f(R,T)$ correction enhances the EH screening effect, while for $\beta>0$, it diminishes it. The $\beta$-dependence introduces a tunable modification to the spacetime curvature.
\par Here we have to regard the choice of parameters. Some articles argue that under the constraints of Solar System PPN measurements \cite{Shabani:2014xvi,Singh:2014bha}, $|2\beta| \leq 2.9 \times 10^{-4}$. In this paper we still adopt the parameter range $|2\beta| \leq 2$ for $f(R,T)$ in cosmology and black hole studies \cite{Jamil:2011ptc,Salehi:2015ira,Santos:2023fgd,Hazarika:2024cji,AraujoFilho:2025hnf}. Additionally, for the EH parameter $a$, we choose $0 \leq a \leq 10$ to remain consistent with Ref. \cite{Yajima:2000kw}.
\par
\begin{figure}
	\begin{center}
		\subfigure[a=0]{\includegraphics[width=5cm]{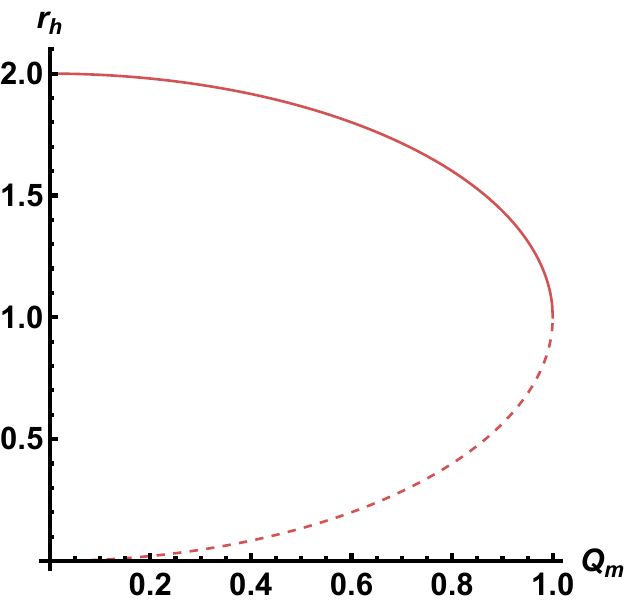}\label{qmb}}
		\subfigure[a=2]{\includegraphics[width=5cm]{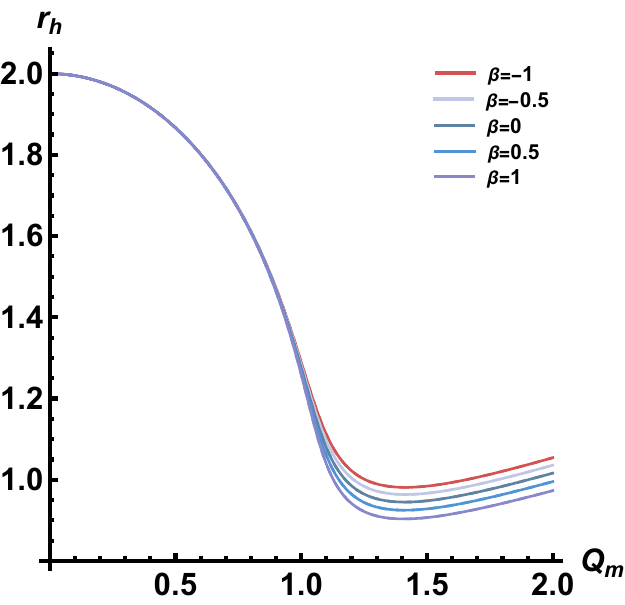}\label{qmrha2}}
		\subfigure[a=10]{\includegraphics[width=5cm]{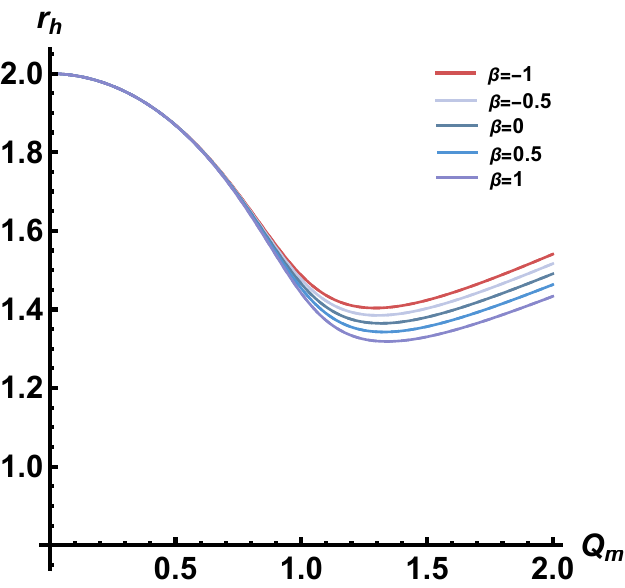}\label{qmrha10}}
	\end{center}
	\caption{The relationship between the outer event horizon $ r_+ $(Solid line), inner event horizon $r_-$(dashed line)  and the charge $ Q_m $ is shown in the figure. With $M = 1$ fixed, $a$ and $\beta$ vary. When $a=0$ the solution reduce to RN black hole.}
	\label{qm}
\end{figure}

 Due to the difficulty in obtaining an analytical solution for the event horizon $ r_h$, we present the numerical solutions between $r_h$ and the magnetic charge $Q_m$ in Fig. \ref{qm}. When the EH parameter $a=0$, the spacetime reverts to the standard magnetically charged RN black hole, and the effect of $\beta$ is screened. At this point, the black hole has an inner horizon $r_- $ and an outer horizon $ r_+ $, with the intersection point corresponding to an extremal black hole. When $a $ is small ($\approx 0$), the inner horizon still exists. When $a$ is large, $r_-$ disappears, and $r_+$ first decreases and then increases as \( Q_m \) increases. For the same $Q_m$, a larger $a$ results in a larger $r_h$ for the black hole. An increase in the coupling strength $\beta$ leads to a decrease in $ r_+ $. Compared to  EEH black holes($\beta=0$ and $\alpha>0$), positive $\beta$ reduces the event horizon radius $r_h$ for fixed $Q_m$, while negative $\beta$ increases it.
\par The Kretschmann scalar reflects the true curvature strength of the gravitational field, which has the following form
\begin{equation}
	\mathcal{K}=R^{\mu\nu\rho\sigma}R_{\mu\nu\rho\sigma},
\end{equation}
and we can obtain
\begin{equation}
	\mathcal{K}=\frac{48}{r^6} - \frac{96Q_m^2}{r^7}+ \frac{56Q_m^4}{r^8} - \frac{8a(2\pi - \beta)Q_m^4(19Q_m^2 - 14r)}{5\pi r^{12}} + \frac{239a^2(\beta - 2\pi)^2 Q_m^8}{50\pi^2 r^{16}}.
\end{equation}
When $r \to 0$, $ \mathcal{K} \to \infty$, it indicates a physical singularity. We observe no divergence of \( \mathcal{K} \) at the event horizon, confirming that the event horizon is a coordinate singularity, not a physical one.
\section{Electrically charged black holes\label{s7}}
\par The BI low energy limit interpretation is adopted for electrically charged black hole
solutions of $f(R, T)-EH$ system \cite{Yajima:2000kw}. The electromagnetic field restricts to an electric charge $Q_e$, the symmetry of the spacetime allows the nonvanishing components
\begin{equation}
	F_{tr}=-F_{rt}=-\frac{Q_e}{r^2},
\end{equation}
and the invariant reads $F=-Q_e^2/2r^4, G=0$. By substitute this to Eq. (\ref{9}) and Eq. (\ref{feq3}) and one can obtain the metric function
\begin{equation}
	A_e(r)=1 - \frac{2M}{r}+ \frac{Q_e^2}{r^2}+ \frac{r^2 \beta}{3\pi}+ \frac{a\beta Q_e^2}{\pi r^2}  + \frac{3Q_e^2 \beta}{2\pi r^2}+ \frac{aQ_e^4}{30r^6} + \frac{a\beta Q_e^4}{60\pi r^6},
	\label{ae}
\end{equation}
we note that an $r^2$ term emerges, and some studies suggest this is an effective cosmological constant $\Lambda_{\text{eff}}$ induced by the $f(R,T)$ correction. When the EH correction $a$ vanishes and one set $\beta=-3\pi \Lambda$, the spacetime reduce to RN-(a)dS black holes \cite{Jafarzade:2024zqq}. If $\beta$ takes a large value (e.g., $|\beta|\sim 1$), it will produce an effective cosmological constant of $\Lambda_\text{eff}\sim 1/\pi$. This value is astronomically large and incompatible with our observed universe ($\Lambda_\text{obs}\sim 10^{-122}$) \cite{Planck:2018vyg}. Therefore, the parameter range $|\beta| \leq 1$ is adopted for illustrative purposes.  
\par Besides $f(R, T)$, other modified gravity mechanisms can also give rise to an $r^2$ term. In $\mathcal{R}^2$ gravity \cite{Jafarzade:2025xcn,Kehagias:2015ata}, $F(R)$ gravity \cite{DeFelice:2010aj,Nojiri:2017ncd,Nojiri:2010wj,Jafarzade:2025nbe}, and higher-order curvature corrections of Gauss–Bonnet \cite{Miskovic:2010ui}, a mathematical structure analogous to the cosmological constant is introduced in the curvature terms of the action, which subsequently generates an $r^2$ term, preventing the spacetime curvature from being asymptotically flat. Some studies have also reported charged solutions yield $\Lambda_{\text{eff}}$ \cite{Rois:2025tfe}. The emergence of the $r^2$ term originates from the $\beta T g_{\mu\nu}$ term in the $f(R,T)$ field equations and the EH electromagnetic energy-momentum tensor $T_{\mu\nu}$. For the electrically charged solution, which employs a radial electric field, the coupling with the trace generates a global effect analogous to a cosmological constant $\Lambda$.
\begin{figure}
	\begin{center}	
		\subfigure[$a=2$]{\includegraphics[width=4cm]{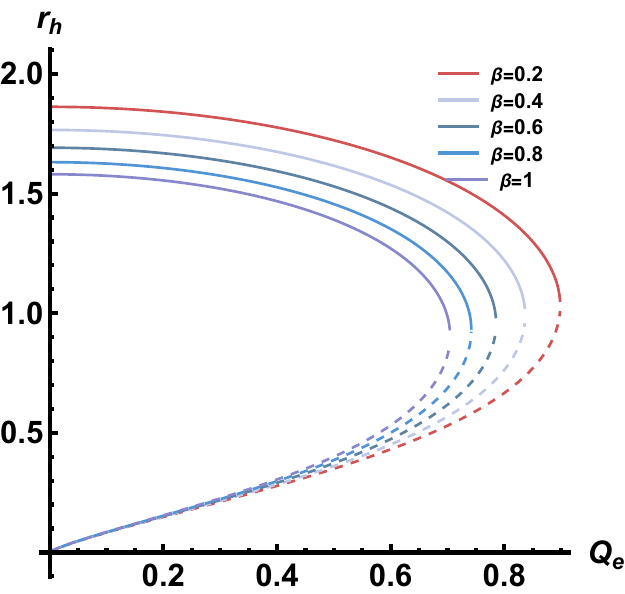}}
		\subfigure[$a=10$]{\includegraphics[width=4cm]{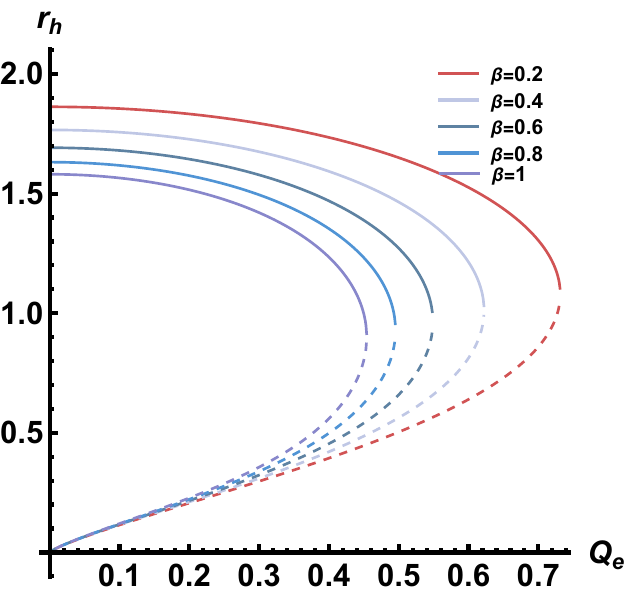}}
		\subfigure[$a=2, \beta=-0.2$]{\includegraphics[width=4cm]{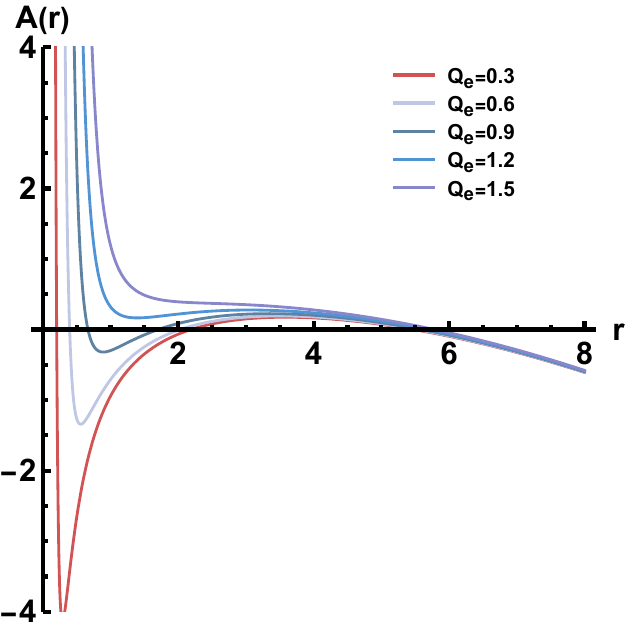}}
		\subfigure[$a=2, \beta=-0.5$]{\includegraphics[width=4cm]{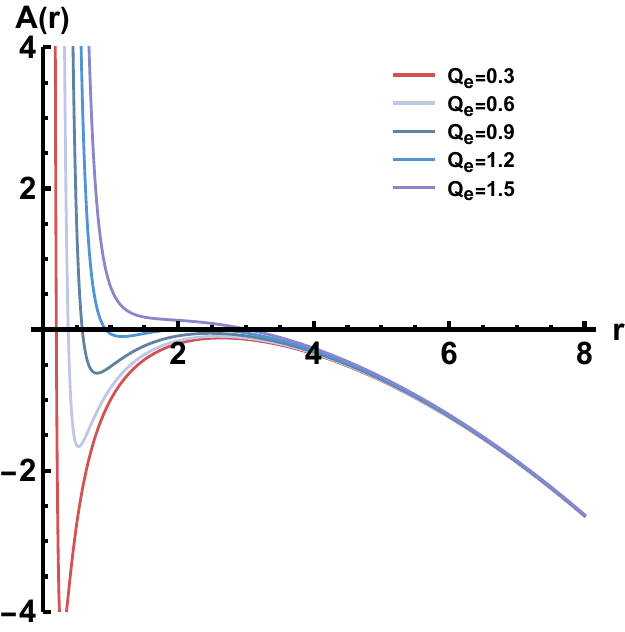}}
	\end{center}
	\caption{The relationship between the outer event horizon $ r_+ $(Solid line), inner event horizon $r_-$(dashed line)  and the charge $Q_e$ are shown in the figure (a) and (b). For the case where $\beta$ is negative (dS spacetime), the plots of A(r) are shown in (c) and (d). All figures have $M =1$ fixed, $a$ and $\beta$ vary. }
	\label{rha}
\end{figure}
\par The relationship between the charge $Q_e $ and the horizon $ r_h $ for different values of $a $ are illustrated in Fig. \ref{rha}. For the case where $\beta > 0$ (AdS spacetime), as $\beta$ increases, both the inner and outer horizons decrease, and the extremal black hole has smaller \( r_h \) and \( Q_e \). An increase in \( a \) makes this effect more pronounced. 
\begin{figure}
	\begin{center}
		\includegraphics[width=0.6\textwidth]{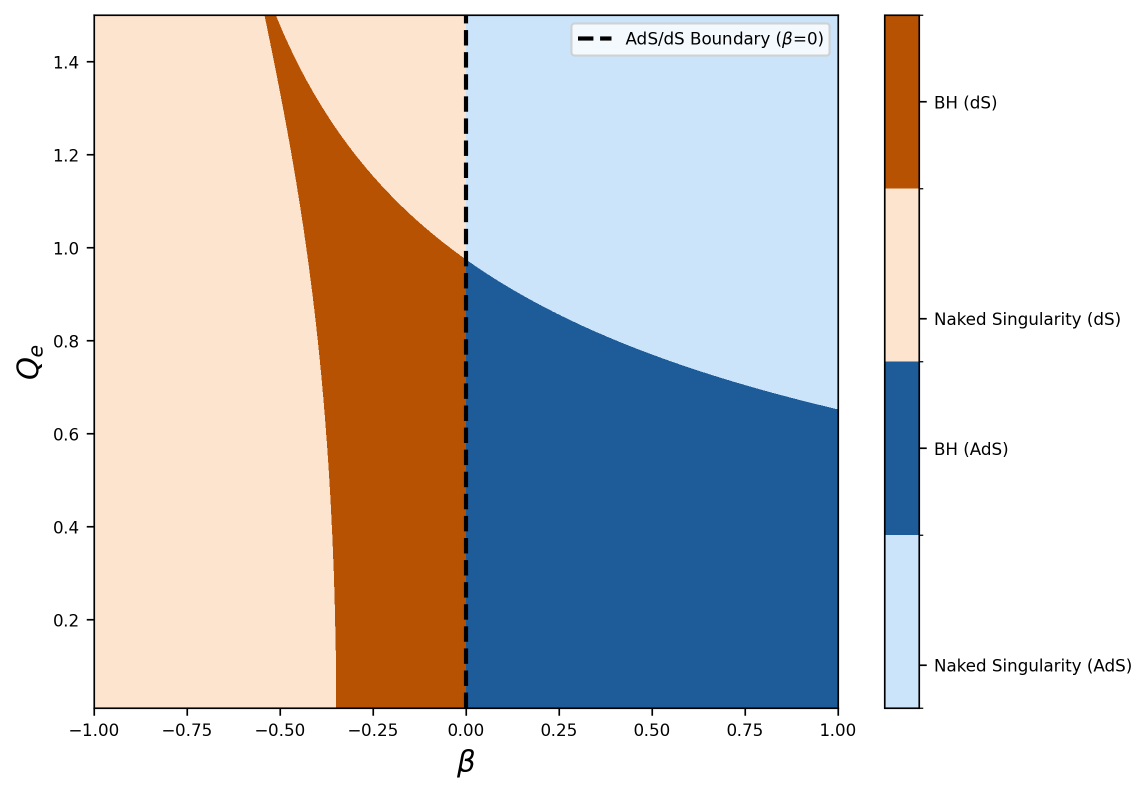}
	\end{center}
	\caption{Phase diagram in the $(Q_e, \beta)$ parameter space for $M=1, a=2$. The central line ($\beta=0$) divides the spacetime into the AdS($\beta>0$) and dS($\beta<0$) regimes. In both regions, darker shades denote the Black Hole (BH) phase (characterized by the existence of event horizons $r_{\pm}$), and lighter shades represent the Naked Singularity (NS) phase (where $r_{\pm}$ are absent or degenerate). The boundary between the BH and NS regions is defined by the Extremal Black Hole locus ($r_{-} = r_{+}$).}
	\label{phase}
\end{figure}
\par For dS spacetime ($\beta <0$, Fig. \ref{rha}(c) and (d)), there exists three distinct horizon structures: (i) Three Horizons: The presence of three horizons, $r_-$, $r_+$, and $r_c$, which is analogous to the standard RN-de Sitter (RN-dS) spacetime. (ii)Two Horizons: The existence of two horizons, $r_e$ and $r_c$, where $r_e$ represents an extremal horizon (a degenerate event horizon). (iii) Single Horizon: The existence of only the cosmological horizon $r_c$, within which lies a naked singularity \cite{Vagenas:2019rai}. One can analyze this by examining the asymptotic form of $A_e(r)$ from Eq. (\ref{ae}) as $r \to 0$. The function is illustrated with $M=1$ and $a=2$ (Without loss of generality), and the phase diagram of $\beta$ and $Q_e$ in Fig. \ref{phase} describes the ranges in the parameter space where these three horizon structures exist. For AdS spacetime ($\beta > 0$), as $\beta$ increases (corresponding to enhanced effective gravitational attraction), the maximum allowable charge $Q_{e, \text{max}}$ decreases, indicating that the curvature correction induced by $\beta$ facilitates the spacetime in reaching the extremal state. In contrast, within the dS region ($\beta < 0$), the naked singularity (NS (dS)) phase emerges at both the low-charge ($Q_e \to 0$) and high-charge limits. Furthermore, the extremal black hole boundary in this domain is severely compressed; as $\beta$ becomes increasingly negative, the parameter space permitting stable black holes (BH (dS)) contracts rapidly.
\par The existence of a naked singularity needs to discuss its causal nature. As $r \to 0$, the $r^{-6}$ term is dominant. Given $Q_e^4 > 0$, the sign depends on $(2 + \beta/\pi)$. We adopt the parameter range $|\beta| \le 1$, the term is strictly positive. Therefore, the $r^{-6}$ term is always positive and leads to $\lim_{r\to 0} A_e(r) \to +\infty$. We now analyze the metric components $g_{tt} = -A_e(r)$ and $g_{rr} = 1/A_e(r)$,
\begin{equation}
	\lim_{r\to 0} g_{tt}(r) = \lim_{r\to 0} -A_e(r) \to -\infty,\qquad
	\lim_{r\to 0} g_{rr}(r) = \lim_{r\to 0} \frac{1}{A_e(r)} \to 0^+.
\end{equation} 
This asymptotic behavior ($g_{tt} \to -\infty, g_{rr} \to 0$) is a timelike singularity. Nonetheless, its existence without a horizon remains a clear violation of the CCH, highlighting a novel and challenging feature of $f(R,T)$-EH gravity.
\par Similarly, we provide the expression for the Kretschmann scalar
\begin{equation}
	\begin{split}
		\mathcal{K}&=\frac{239 a^2 \beta ^2 Q_e^8}{450 \pi ^2 r^{16}}+\frac{478 a^2 \beta Q_e^8}{225 \pi  r^{16}}+\frac{478 a^2 Q_e^8}{225 r^{16}}+\frac{304 a^2 \beta Q_e^6}{15 \pi  r^{12}}+\frac{152 a^2 \beta ^2 Q_e^6}{15 \pi ^2 r^{12}}+\frac{76 a \beta ^2 Q_e^6}{15 \pi ^2 r^{12}}\\&+\frac{304 a \beta  Q_e^6}{15 \pi  r^{12}}+\frac{304 a Q_e^6}{15 r^{12}}-\frac{112 a \beta  Q_e^4}{15 \pi  r^{11}}-\frac{224 a Q_e^4}{15 r^{11}}+\frac{508 a \beta ^2 Q_e^4}{9 \pi ^2 r^8}+\frac{1016 a \beta  Q_e^4}{9 \pi  r^8}\\&+\frac{56 a^2 \beta ^2 Q_e^4}{\pi ^2 r^8}+\frac{14 \beta ^2 Q_e^4}{\pi ^2 r^8}+\frac{56 \beta Q_e^4}{\pi  r^8}+\frac{56 Q_e^4}{r^8}-\frac{96 a \beta  Q_e^2}{\pi  r^7}-\frac{48 \beta  Q_e^2}{\pi  r^7}-\frac{96 Q_e^2}{r^7}+\frac{48}{r^6}+\frac{8 \beta ^2}{3 \pi ^2}.
	\end{split}
\end{equation}
\par The Kretschmann scalar indicates a coordinate singularity at the horizon, which can be eliminated through a coordinate transformation, while $r = 0$ is a physical singularity. Unlike the magnetic charge solution, due to the presence of the “effective cosmological constant" $\Lambda_{\text{eff}}$, the spacetime curvature does not vanish at infinity but approaches a constant value of \( 8\beta^2/(3\pi) \). When \( \beta \) is negative, there exists a cosmological horizon $r_c$, corresponding to a dS spacetime with positive curvature. When $\beta$ is positive it corresponds to an AdS spacetime.
\section{Black holes with dyonic EH electrodynamics\label{s4}}
\par In Sections III and IV, we discuss the $f(R,T)$ black hole solutions with magnetic and electric charges within the framework of EH as a low-energy limit of BI theory. For the dyonic case (where both electric and magnetic charges are present), as the electric and magnetic fields approach the critical limits, QED one-loop corrections become especially pronounced, including photon–photon scattering and vacuum birefringence \cite{Ruffini:1971bza,Ruffini:2013hia,Novello:1999pg,Jafarzade:2025byr}. This interpretation is consistent with quantum corrections in particle physics and is suitable for describing dyonic systems where electromagnetic interactions are dominant. Furthermore, the shadow features observed in EHT data may be influenced by birefringence, providing further justification for employing the QED interpretation in the dyonic analysis.
\par In the dyonic case, the gauge potential has the form $A_\mu=(A_t(r),0,0,Q_m \cos \theta)$, where $A_t(r)$ represents the contribution of the electric field to the gauge potential, reads \cite{Hawking:1995ap,Magos:2023nnb}
\begin{equation}
	A_t(r) =- \frac{Q_e}{r} \left( 1 - \frac{2\alpha}{225\pi} \frac{Q_e^2}{r^4E_c} - \frac{\alpha}{45\pi} \frac{Q_m^2}{r^4E_c} \right),
\end{equation}
where $\alpha$ is the fine structure constant and $E_c\equiv m_e^2c^3/e\hbar$. For simplicity we set $\eta=2\alpha/225\pi E_c$ in our further calculation. Thus, the nonzero components of the Faraday tensor are
\begin{equation}
	\begin{split}
		F_{tr}&=-F_{rt}=-\frac{Q_e}{r^2}\left(\frac{1}{r^2}-\frac{10\eta Q_e^2+25\eta Q_m^2}{2r^2}\right),\\
		F_{\theta\phi}&=-F_{\phi\theta}=Q_m\sin\theta,
	\end{split}
\end{equation}
and the invariant of electromagnetic field reads
\begin{equation}
	F=\frac{1}{4}F^{\mu\nu}F_{\mu\nu}=\frac{Q_m^2}{2r^4}-\left(\frac{ Q_e\left(5 \eta  \left(2 Q_e^2+5 Q_m^2\right)-2 r^4\right)}{8 r^6}\right)^2,
\end{equation}
\begin{equation}
	G=\frac{1}{4}F_{\mu\nu}*F^{\mu\nu}=-\frac{Q_m}{r^2}\frac{ Q_e\left(5 \eta  \left(2 Q_e^2+5 Q_m^2\right)-2 r^4\right)}{2 r^6}.
\end{equation}
By substituting this to Eq. (\ref{9}) and Eq. (\ref{feq3}) , we have the analytical solution of the dyonic case
\begin{equation}
	\begin{split}
		&A_{dy}(r)=1-\frac{2M}{r}+\frac{Q_e^2+Q_m^2}{r^2}- \frac{\eta Q_e^2\epsilon}{3r^6}+ \frac{a(Q_e^4 + 9Q_e^2Q_m^2 - 3Q_m^4)}{30r^6}\\&- \frac{a\beta(Q_e^2 + 7Q_e^2Q_m^2 - 15Q_m^4)}{60\pi r^6}   + \frac{5\eta\epsilon}{36r^{10}}+\frac{5\pi\eta^2 Q_e^2\epsilon^2}{36r^{10}}+a\eta  Q_e^2 \epsilon \frac{2 \beta  Q_e^2-4 \pi Q_e^2+7 \beta  Q_m^2-18 \pi  Q_m^2}{36 \pi  r^{10}}\\&-25 a \eta ^2 \epsilon^2 \frac{6 \beta  Q_e^2-12 \pi  Q_e^2+7 \beta  Q_m^2-18 \pi  Q_m^2}{1456 \pi  r^{14}}-\frac{125 a (2 \pi -\beta ) \eta ^3 Q_e^4 \epsilon^3}{1224 \pi  r^{18}}+\frac{625 a (2 \pi -\beta ) \eta ^4 Q_e^4\epsilon^4}{14784 \pi  r^{22}},
	\end{split}
\end{equation}
\begin{figure}
\begin{center}
	\subfigure[ $\beta=-1$]{\includegraphics[width=5cm]{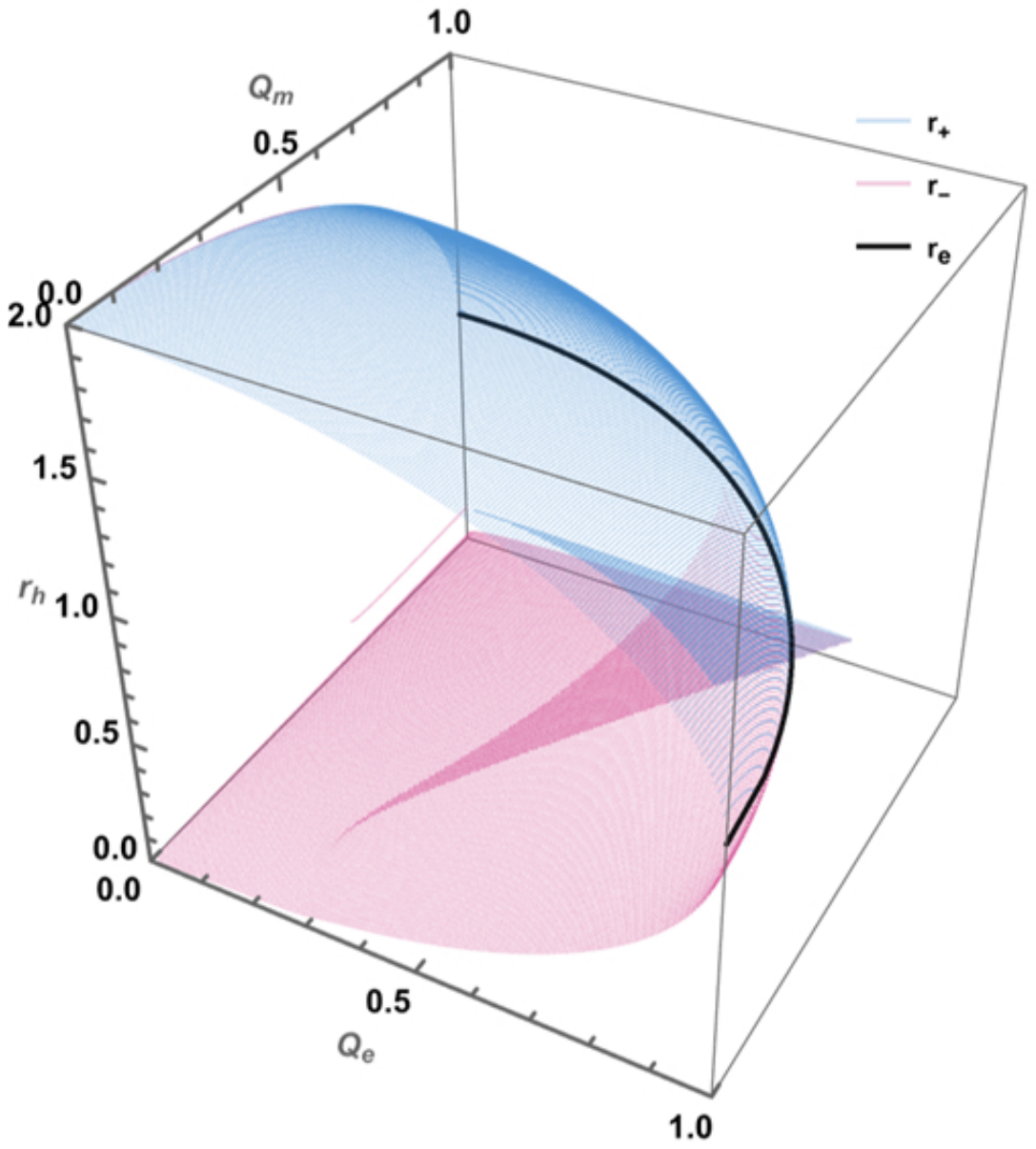}\label{qebqe03}}
	\subfigure[$\beta=1$]{\includegraphics[width=5cm]{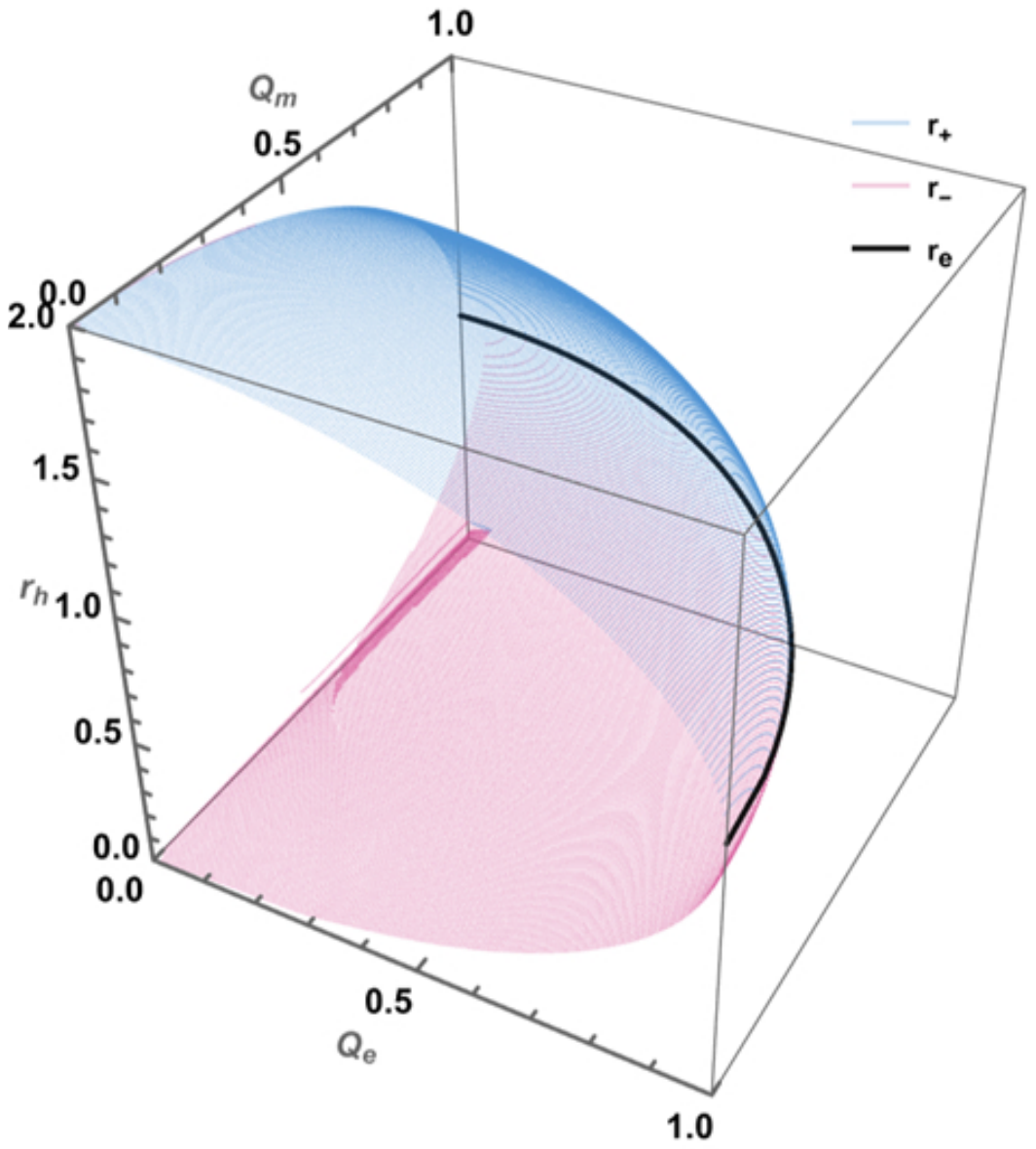}\label{qebqe05}}
\end{center}
\caption{The relationship between the outer event horizon $ r_+ $(blue), inner event horizon $r_-$(pink), extremal black hole horizon $r_e$(black curve)  and the charge $ Q_e, Q_m$ is shown in the figure. With \( M = 1 \) fixed, $a=0.001768$ and $\beta$ vary. }
\label{qeqm}
\end{figure}
where $\epsilon=2Q_e^2+5Q_m^2$, and the EH parameter $a=0.001768$. We note that when the coupling strength parameters \(\eta\) and \(a\) vanishes, the spacetime reverts to a Reissner-Nordstr\"{o}m (RN)-like solution with magnetic charge \cite{Wald:1974np}. Note that in the corrections of $1/r^6$ and higher order, $Q_m$ and $Q_e$ no longer appear symmetrically. Consequently, the spacetime electromagnetic field near the black hole’s horizon exhibits significant deviations from the classical linear Maxwell electromagnetic theory. This correction leads to notable observational effects in the strong-field regime, which will be discussed in Section \ref{s6}.
\par We present the relationship between $ r_h $ and the dual charges in Fig. \ref{qeqm}. The blue surface represents the outer horizon $ r_+ $, the pink surface represents the inner horizon $ r_- $, and the black curve at their intersection represents the extremal black hole. Increasing magnetic and electric charges enlarges the inner horizon and shrinks the outer horizon, trending toward an extreme black hole. When $\beta$ is positive, the overall variation trend of the horizon is consistent with that of the $ RN $ black hole with electric and magnetic charge. However, when $\beta$ is negative and the $Q_e,Q_m$ are close to each other, a larger $ r_- $ is obtained.

\par Similar to the magnetic charge EH field calculation, we provide the Kretschmann curvature scalar $\mathcal{K}$ expression for the dyonic case in the appendix \ref{ap}. Noting that the event horizon \( r_h \) remains a coordinate singularity, with a physical singularity only at \( r \to 0 \). Both electric and magnetic charges significantly affect spacetime curvature.
\section{energy conditions\label{s5}}
\par In this section we briefly review the energy condition in general relativity and how the energy condition constrains the black hole metric. The energy conditions are expected to impose restrictions on the stress-energy momentum tensor $T_{\mu\nu}$ of matter fields in a physically
reasonable model \cite{Guo:2022ghl}. The origin of energy conditions is the Raychaudhuri equation along with the requirement that gravity is attractive,
\begin{equation}
	\frac{d\theta}{d\tau} = -\frac{1}{3}\theta^2 - \sigma_{\mu\nu}\sigma^{\mu\nu} + \omega_{\mu\nu}\omega^{\mu\nu} - R_{\mu\nu}u^\mu u^\nu ,
\end{equation}
where $\theta$, $\sigma_{\mu\nu}$ and $\omega_{\mu\nu}$ are the expansion, shear and rotation associated to the vector field $u_\mu$  congruent to timelike geodesics. For any timelike 4-vectors $t^\mu$,  the weak energy condition (WEC) and the strong energy condition (SEC) requires that the energy-momentum tensor satisfies \cite{Santos:2007bs}
\begin{equation}
	T_{\mu\nu}^{\text{eff}}t^\mu t^\nu \geq 0 \qquad\text{WEC},\\
	\label{20}
\end{equation}
\begin{equation}
	(T_{\mu\nu}^{\text{eff}}-\frac{1}{2}T^{\text{eff}}g_{\mu\nu})t^\mu t^\nu \geq 0 \qquad\text{SEC},\\
	\label{21}
\end{equation}
the effective stress-energy momentum tensor $T_{\mu\nu}^{\text{eff}}$ and its trace $T^{\text{eff}}$ of magnetic charged and dyonic charged EH $f(R,T)$  spacetime have the form \cite{Harko:2011kv}
\begin{equation}
	\begin{split}
		T^{\text{eff}}_{\mu\nu} =& T_{\mu\nu} + \frac{f_T}{8\pi}(T_{\mu\nu} + \Theta_{\mu\nu})\\
		=&\frac{g_{\mu\nu} }{4\pi} \left( -F + a F^2 + \frac{7}{4} a G^2 \right) - \frac{1}{4\pi} \left( -1 + 2aF \right) F_{\mu \rho} F^{\rho}_{\nu}+\frac{\beta a}{4\pi^2}FF_{\mu\nu}F_\nu^\rho,\\
		T^{\text{eff}}=&g^{\mu\nu}T_{\mu\nu}^{\text{eff}},
	\end{split}	
	\label{22}
\end{equation}
where $f_T=\partial f(R,T)/\partial T=2\beta$. We choose a static observer with timelike 4-vector $t_\mu=(1/\sqrt{A(r)}, 0, 0, 0)$ which satisfies $g_{\mu\nu}t^\mu t^\nu=-1$, and by substituting Eq. (\ref{9}) and  Eq. (\ref{22}) to  Eq. (\ref{20}) and Eq. (\ref{21}), we notice that due to $Q_m>0$ and $Q_e>0$ the spacetime satisfies the WEC outside the event horizon in both magnetic and dyonic charged case. 
\par We note that the spacetime of a magnetic charge black hole violates the SEC and WEC near the singularity, with the boundary of the WEC lying within that of the SEC. The electrically charged AdS black hole ($\beta > 0$) globally satisfies both the SEC and WEC, while the electrically charged dS black hole ($\beta < 0$) globally satisfies the WEC but only satisfies the SEC in the region near the horizon. For the dyonic black hole, both the SEC and WEC are satisfied outside the inner horizon \cite{Guo:2022ghl,Cvetic:2016bxi}. 
\section{Geodesics and ISCO of black holes in $f(R,T)$ gravity coupled with EH field\label{s6}}
\par Due to the nonlinear electrodynamics effects, photons propagate along null geodesics in the effective metric $G^{\mu\nu}$ rather than the background metric, which reads \cite{Novello:1999pg,AraujoFilho:2024lsi,AraujoFilho:2024xhm}
\begin{equation}
	G^{\mu\nu}=\mathcal{L}_{FF} F^{\mu}_{\alpha}F^{\alpha\nu}-\mathcal{L}_F g^{\mu\nu},
	\label{eff}
\end{equation}
and we can set ansatz of effective metric for spacetime in $f(R,T)$ gravity coupled with EH magnetic monopole and dyonic charge 
\begin{equation}
		ds^2=h(r)(-w(r)dt^2+\frac{1}{w(r)}dr^2+u(r)(d\theta^2+\sin^2\theta d\phi^2)),
		\label{effg}
\end{equation}
and we consider the geodesic equation
\begin{equation}
	\frac{d^2 x^\mu}{d\lambda^2} + \Gamma^\mu_{\rho\sigma} \frac{dx^\rho}{d\lambda} \frac{dx^\sigma}{d\lambda} = 0,
\end{equation}
where $\lambda$ is the arbitrary affine parameter, and $\Gamma^\mu_{\rho\sigma}$ is the Christoffel symbol. We set $p^\mu\equiv dx^\mu/d\lambda$  to describe the 4-momentum of the photon. The metric $G_{\mu\nu}$ does not explicitly depend on coordinates $t$ and $\phi$, there exists two conserved quantities, energy $E$ and the angular momentum $L$. The symmetry provides two Killing vector
\begin{equation}
	\begin{split}
		K_\mu&=(-h(r)w(r),0,0,0),\\
		R_\mu&=(0,0,0,h(r)u(r)\sin^2\theta),
	\end{split}	
\end{equation}
and we could obtain the energy $E$ and the angular momentum $L$ for $\theta= \pi/2$  along the geodesics \cite{Liang:2024xif}
\begin{equation}
	E=-K_\mu p^\mu=h(r)w(r)p^t,\quad L=R_\mu p^\mu=h(r)u(r)p^\phi,
	\label{el}
\end{equation}
substituting Eq. (\ref{effg}) and Eq. (\ref{el}) to the Lagrangian of photons $\mathcal{L}=\frac{1}{2}g_{\mu\nu}p^\mu p^\nu=0$, we can derive
\begin{equation}
	-\frac{E^2}{h(r)w(r)} + \frac{h(r)}{w(r)} (p^r)^2 + \frac{L^2}{h(r)u(r)} = 0.
\end{equation}
We define the impact parameter $b\equiv |L|/E$ and setting the affine parameter $\lambda\rightarrow\lambda/L$, one can derive the equation of photon motion
\begin{equation}
	\frac{dr}{d\phi} = \pm u(r) \sqrt{\frac{1}{b^2} - \frac{w(r)}{u(r)}},
\end{equation}
where $\pm$ represents the counterclockwise and clockwise geodesics of the photon. 
\par One can set $V(r) = w(r)/u(r)$ as the effective potential of photons and notice that photons can circle the black hole many times in unstable orbits, called the photon sphere, and it satisfies
\begin{align}
	b_{\text{ph}}=\frac{1}{\sqrt{V(r_{\text{ph}})}},\qquad \frac{dV(r_{\text{ph}})}{d r}=0,\qquad\frac{d^2V(r_{\text{ph}})}{d r^2}\leq0,
\end{align}
here $r_{\text{ph}}$ is the radius of photon sphere. Since photon sphere is the innermost unstable circular orbit of photon, one can use the $r_{\text{ph}}$ to describe the radius of the black hole shadow.
\par As mentioned in Refs. \cite{Harko:2011kv,Pretel:2022qng}, due to the violation of the covariant conservation of the energy-momentum tensor in $f(R,T)$ gravity, i.e.,
\begin{equation}
	\nabla^\mu T_{\mu\nu} = \frac{f_T(R,T)}{8\pi - f_T(R,T)} \left[ \left( T_{\mu\nu} + \Theta_{\mu\nu} \right) \nabla^\mu \ln f_T(R,T) + \nabla^\mu \Theta_{\mu\nu}-\frac{1}{2}g_{\mu\nu}\nabla^\mu T \right]\neq0,
\end{equation}
uncharged massive particles do not follow timelike geodesics, and the equation of motion should be modified by extra force $f^\alpha$ as follows, which is considered as the projection of $\nabla^\mu T_{\mu\nu}$ in the direction perpendicular to \( u^\mu \), defined by the projection operator $h_{\mu\alpha}$ \cite{Bertolami:2007gv,Harko:2010mv,Haghani:2013oma,Xu:2019sbp}
\begin{equation}
	\begin{split}
		&\frac{d^2 x^\alpha}{d\tau^2} + \Gamma^\alpha_{\mu\nu} \frac{dx^\mu}{d\tau} \frac{dx^\nu}{d\tau} = f^\alpha, \\
		&f^\alpha = \frac{\beta}{4\pi +\beta} \left(h^{\alpha\mu} \nabla_\mu \ln \beta + \frac{1}{2} h^{\alpha\mu} \nabla_\mu T \right),
	\end{split}
\end{equation}
where the $h_{\mu\alpha}=g_{\mu\alpha} -u_\mu u_\alpha$ and satisfies $h_{\mu\alpha}u^\mu=0$. It is noted that when \(\beta = 0\), the energy-momentum tensor is covariantly conserved, reverting to Einstein's gravity, and simultaneously, the extra force \(f^\alpha = 0\), causing the particle's motion to revert to timelike geodesics. If we simply consider an uncharged test particle (e.g. a dust particle) we can still use the definition of the effective potential in Einstein gravity
\begin{equation}
	V_{\text{eff}}\equiv w(r)(1+\frac{L_m}{r^2}),
\end{equation}
where $L_m$ is the angular momentum of test particles. The innermost stable circular orbit (ISCO) is the closest distance at which particles around a black hole can maintain a stable circular orbit \cite{Jefremov:2015gza}, and satisfies the following conditions
\begin{equation}
	\left.\frac{dV_{\text{eff}}(r)}{dr} \right|_{r=r_{\text{ISCO}}}=0, \quad \left.\frac{d^2V_{\text{eff}}(r)}{dr^2}\right|_{r=r_{\text{ISCO}}}>0.
\end{equation}
\par We could use the orbital plane azimuthal angle $f$ to define the number of orbits $n = f/2\pi$, which depends on the impact parameter b \cite{Gralla:2019xty}. According to the different values of n, orbits can be categorized into distinct types: direct emission occurs when $n<3/4$, lensing ring is formed when $3/4<n<5/4$, photon ring is established when $n>5/4$. 
\par We have provided a general analysis of photon geodesics and the ISCO, and we then will separately study the effective metrics for magnetic, electric, and dyonic solutions, integrate the photon trajectories to plot the geodesic images, and provide numerical solutions for $r_{\text{ISCO}}$.
\subsection{magnetically charged case}
By solving the Eq. (\ref{eff}), we have the magnetic charged effective metric
\begin{equation}
		w_m(r)=A_m(r),\quad h_m(r)=\frac{\pi  r^4 }{4 r^4-4 a Q_m^2},\quad u_m(r)=\frac{8r^2(aQ_m^2-r^4)}{8 r^4-9 a Q_m^2}.
\end{equation}
\begin{table}
	\setlength{\tabcolsep}{3mm}
	\begin{center}
		\begin{tabular}[b]{c|cccccc}
			\midrule[2pt]
			$Q_m$ & 0(Sch) & 0.1 & 0.4 & 0.6 & 0.8 & 1 \\
			\toprule[1pt]
			$r_{\text{ph}}$ & 3  & 2.99338 & 2.89054 & 2.74121 & 2.49996 & 2.10741 \\
			$r_{\text{ISCO}}$ &6  & 5.98497 & 5.75284 & 5.42022 & 4.89279 & 4.01471 \\
			\midrule[1pt]
		\end{tabular}

		\begin{tabular}[b]{c|cccccc}
			\midrule[1pt]
			$\beta(a=2)$ & -1 & -0.7 & -0.3 &0.3 & 0.7& 1  \\
			\toprule[1pt]
			$r_{\text{ph}}$ & 2.74138 & 2.74134 & 2.74130 & 2.74123 &2.74118& 2.74114  \\
			$r_{\text{ISCO}}$ & 5.42031 & 5.42029 & 5.42027 & 5.42023 &5.42020& 5.42018 \\
			\midrule[1pt]
			$\beta(a=5)$ & -1 & -0.7 & -0.3 &0.3 & 0.7& 1  \\
			\toprule[1pt]
			$r_{\text{ph}}$ & 2.55207 & 2.55167 & 2.55114 & 2.55034 &2.54981& 2.54941  \\
			$r_{\text{ISCO}}$ & 4.89700 & 4.89674 & 4.89640 & 4.89589 &4.89555& 4.89529 \\
			\midrule[1pt]		
		\end{tabular}
	\end{center}
	\centering
	\caption{Data of  $r_{\text{ph}}$ and $r_{\text{ISCO}}$ of magnetically charged black holes in $f(R,T)$ gravity coupled with EH. Upper table:  for different $Q_m$  with fixed $M=1, \beta=0.4, a=2$. When $a\rightarrow0$, the black hole reduce to the RN-like magnetic charged type, when $Q_m=0$ the black hole reduce to Schwarzschild. Middle table:  for different $\beta$ and fixed $Q_m=0.8, a=2$. When $\beta=0$ the black holes reduce to Einstein-Euler-Heisenberg black holes. Lower table: for different $\beta$ and fixed $Q_m=0.8, a=5$.}
	\label{t1}
\end{table}
\begin{figure}
	\begin{center}
		\subfigure[$\beta=-0.6,Q_m=0.7$]{\includegraphics[width=5cm]{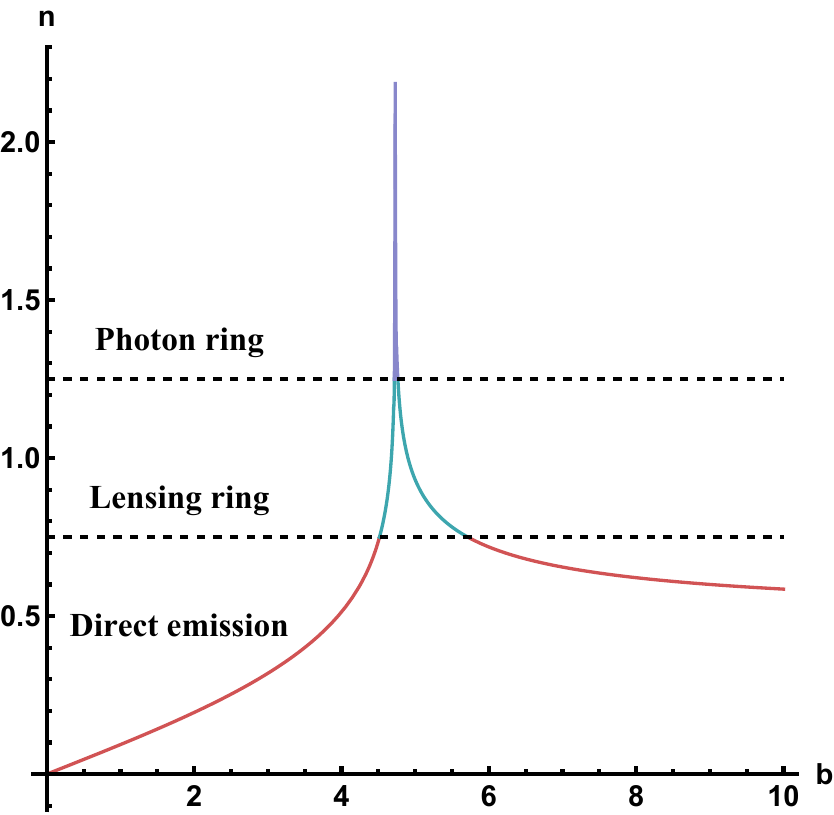}}
		\subfigure[$\beta=0.6,Q_m=0.7$]{\includegraphics[width=5cm]{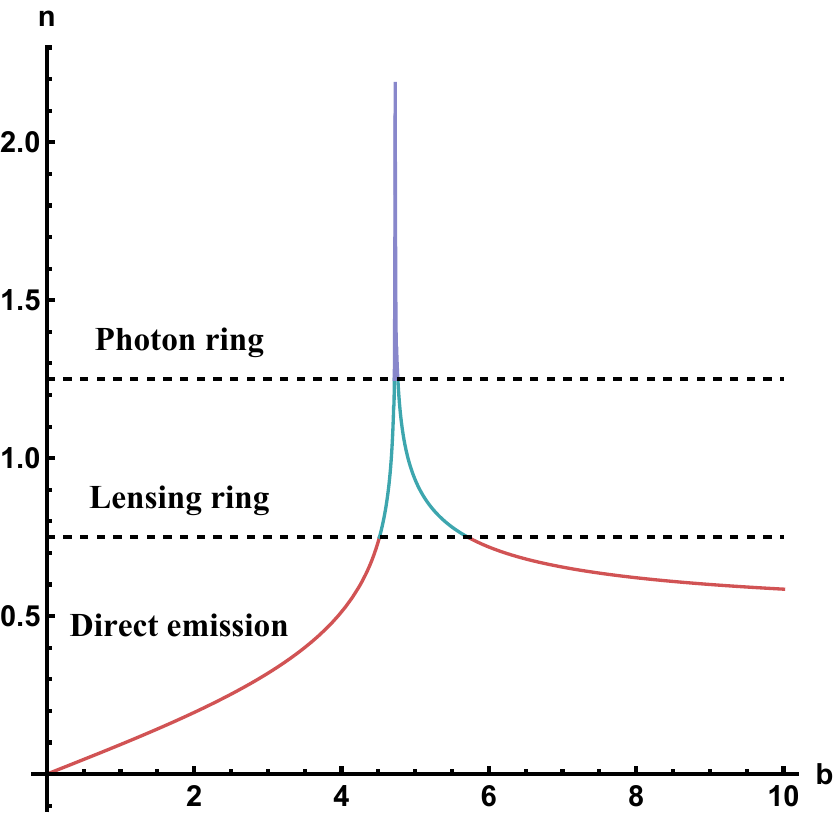}}
		\subfigure[$\beta=0.6,Q_m=0.4$]{\includegraphics[width=5cm]{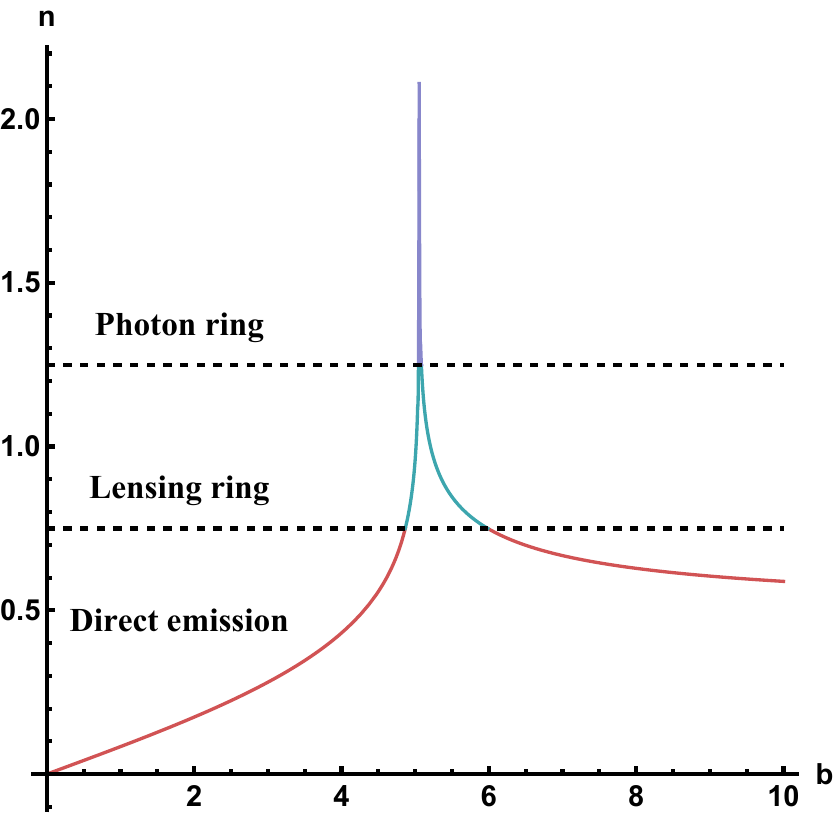}}
		\subfigure[$\beta=-0.6,Q_m=0.7$]{\includegraphics[width=5cm]{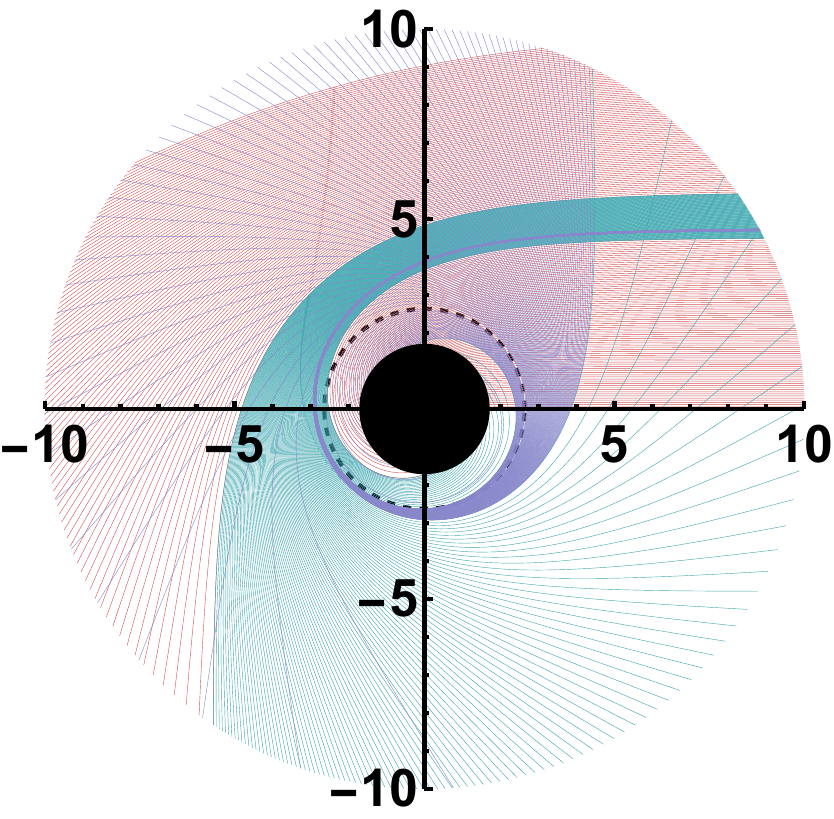}}	
		\subfigure[$\beta=0.6,Q_m=0.7$]{\includegraphics[width=5cm]{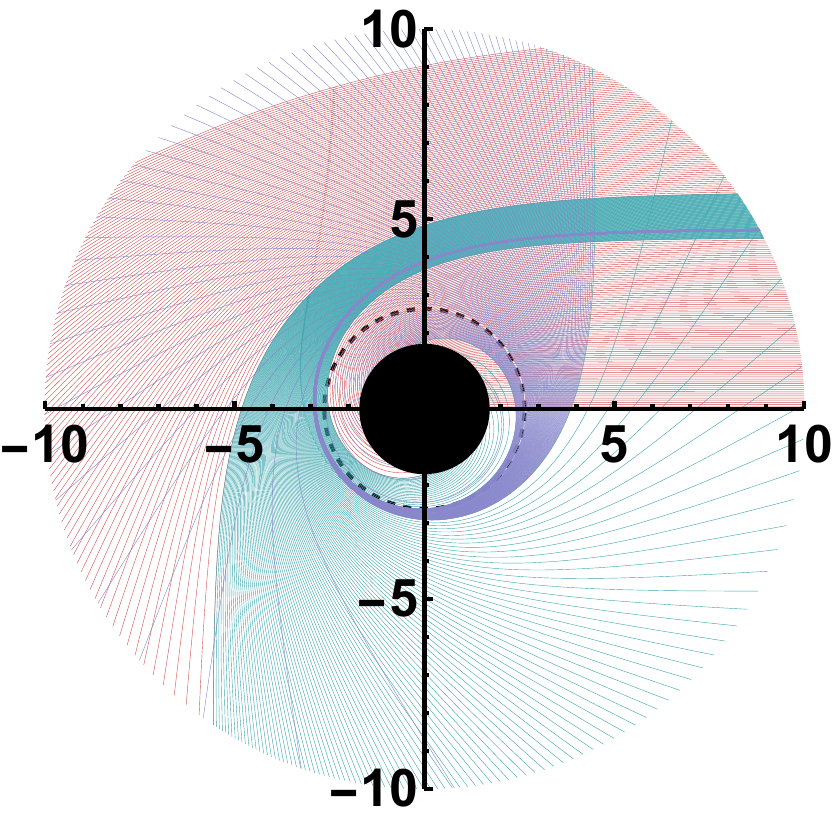}}
		\subfigure[$\beta=0.6,Q_m=0.4$]{\includegraphics[width=5cm]{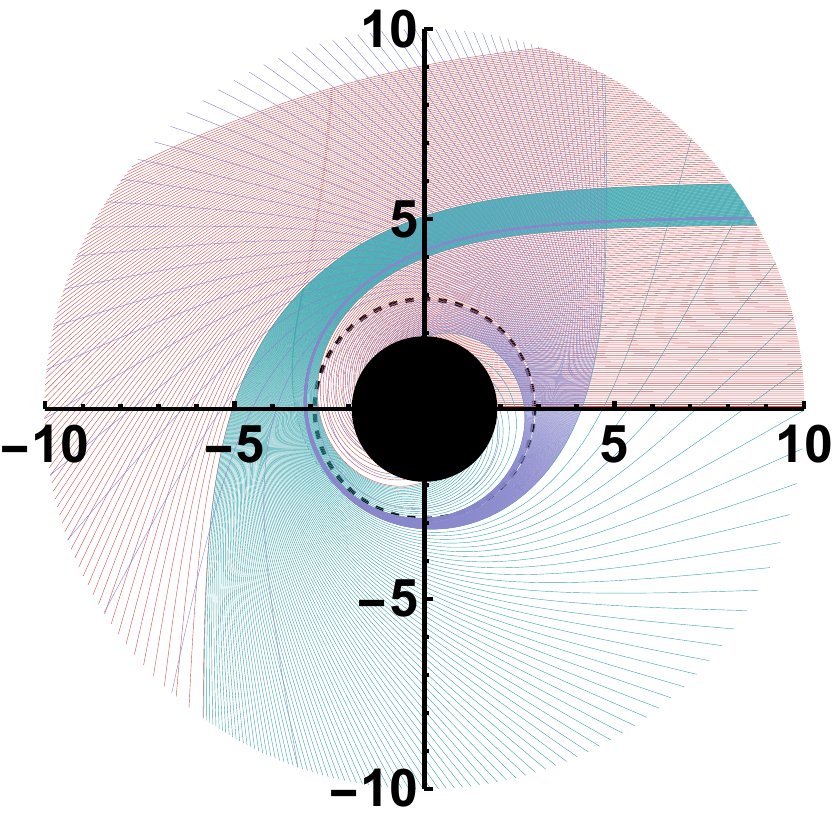}}
	\end{center}
	\caption{Geodesics structure of magnetically charged black hole in $f(R,T)$ gravity coupled with EH field. Upper line are the number of orbits $n$ with respect to impact parameter $b$. Lower line are the geodesics figure. All fixed $a=2$ and $M=1$.}
	\label{geo1}
\end{figure}
\par Since it is not easy to obtain the analytical solution of $r_{\text{ph}}, r_{\text{ISCO}}$, here we list the numerical solution in Table. \ref{t1}. We note that as the $Q_m$  increases, $r_{ph}$, and $ r_{\text{ISCO}}$ significantly decrease. With fixed electric charge, as $\beta$ increases, the photon sphere radius $r_{ph}$ and $r_{\text{ISCO}}$ both decrease. From the comparison between the middle and lower tables, we find that when the change in $\beta$ is the same, the larger the value of a, the more significant the changes in the shadow radius and ISCO radius. This indicates that when the EH correction is significant, the influence of modified gravity on observational features also increases. This correction results in a theoretical deviation in the shadow radius, which may have observable effects.
\par The geodesics of photon  and the relationship of loop numbers $n$ with respect to impact parameter $b$ are plotted in Fig. \ref{geo1}. Increasing magnetic charge causes photons with lower impact parameter \( b \) to achieve a higher $n$, with a significant increase in the proportion of photons in the photon ring and lensing ring, making the overall gravitational lensing effect more pronounced. Increasing the coupling constant \( \beta \) enlarges the horizon radius, shrinks the photon sphere, and increases spacetime curvature near the photon sphere, resulting in photons in the photon ring having a higher number of orbits \( n \).
\subsection{electrically charged case}
\par One can still provide the effective metric reads
\begin{equation}
		w_e(r)=A_e(r),\quad h_e(r)=\frac{2 \pi  r^4}{9 a Q_e^2+8 r^4},\quad u_e(r)=\frac{r^2(9 a Q_e^2+8 r^4)}{8aQ_e^2+8r^4},
\end{equation}
based on the effective metric and the previously mentioned definitions of the photon sphere radius(or shadow radius) $r_{\text{ph}}$ and ISCO, we provide the reference values in Table \ref{t3}.
\begin{table}
	\setlength{\tabcolsep}{3mm}
	\begin{center}
		\begin{tabular}[b]{c|cccccc}
			\midrule[2pt]
			$Q_e$ & 0 & 0.2 & 0.4 & 0.6 & 0.8 & 0.9 \\
			\toprule[1pt]
			$r_{\text{ph}}$ & 3 & 2.96076 & 2.83616 & 2.59747 & 2.21254 & - \\
			$r_{\text{ISCO}}$ & 3.89065 & 3.84351 & 3.69403 & 3.40893 & 2.85593 & 2.11178 \\
			\midrule[1pt]
		\end{tabular}
		\begin{tabular}[b]{c|cccccc}
			\midrule[1pt]
			$\beta$ & -1 & -0.7 & -0.3 &0.3 & 0.7& 1  \\
			\toprule[1pt]
			$r_{\text{ph}}$ & 2.73071 & 2.32297 & 2.80497 & 2.67075 &2.57193& 2.49135  \\
			$r_{\text{ISCO}}$ & 3.57214 & 3.43032 & 4.05537 & 3.60397 &3.36013& 3.22798 \\
			\midrule[2pt]
		\end{tabular}
	\end{center}
	\centering
	\caption{Data of $r_{\text{ph}}$ and $r_{\text{ISCO}}$ of black holes in $f(R,T)$ gravity coupled with EH electrical charge. Upper table:  for different $Q_e$  with fixed $M=1,a=2, \beta=0.6$.  Lower table:  for different $\beta$ and fixed $M=1,a=2, Q_e=0.6$. When $\beta\rightarrow0$, the black hole reduce to the EH electrical charged black hole.}
	\label{t3}
\end{table}
We can conclude that for $\beta>0$(AdS spacetime), as $Q_e$ and $\beta$ increases, both the $r_{\text{ph}}$ decrease significantly. For $\beta<0$(dS spacetime), when there exists singularity (e.g. $\beta=1,0.7$), the photon sphere is outside the cosmological horizon $r_c$ and it means there is no photon sphere in this spacetime; when there remains $r_-$, $r_+$ and $r_c$, the $r_{\text{ph}}$ is larger than that of the singularity.
\par We present the orbit number $ n $ and geodesics of a dS solution in Fig. \ref{gph1}, \ref{gph11} and \ref{g1}, \ref{g11}. Due to the existence of the cosmological horizon, we only consider photons with $ r < r_c $, as photons beyond this are not causally connected to our universe. We note that an increase in $|\beta|$ significantly reduces the cosmological horizon while markedly increasing the impact parameters of the photon ring and the lensing ring. Within the range of impact parameters ($0 < b < 10$), most directly emission photons will fall into the black hole.
\begin{figure}
	\begin{center}
		\includegraphics[width=0.6\textwidth]{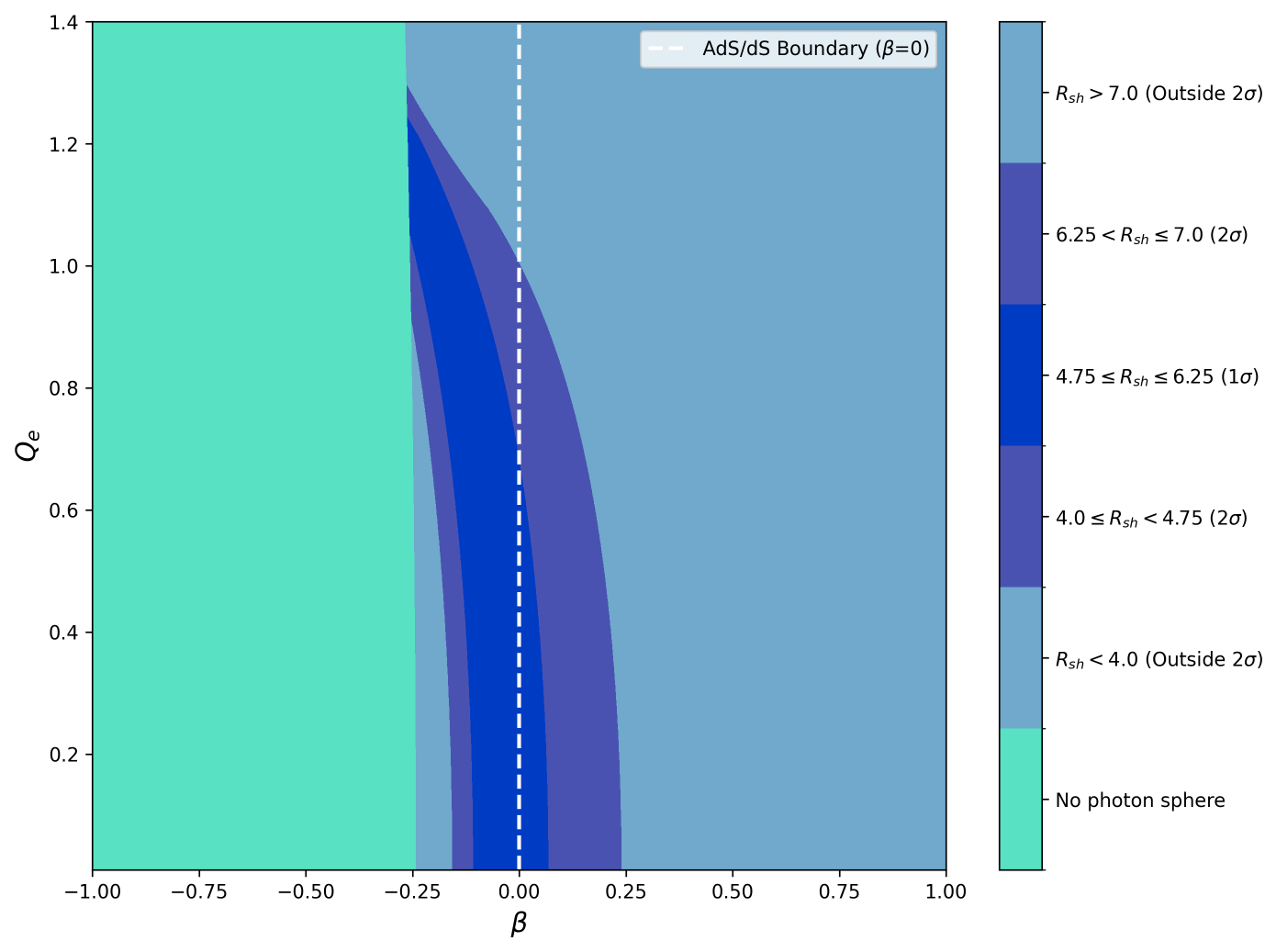}
	\end{center}
	\caption{Constraints of the $(Q_e, \beta)$ for $M=1, a=2$ in EHT M87* observation. The central line ($\beta=0$) divides the spacetime into the AdS($\beta>0$) and dS($\beta<0$) regimes.}
	\label{phase2}
\end{figure}
\par Due to the presence of the $\Lambda_{\text{eff}}$, the spacetime is no longer asymptotically flat. We hypothesize that $\beta$ is an effective parameter that may differ at various energy scales, and we use astrophysical observations (rather than cosmological ones) to independently constrain its value at the black hole scale. We constrain our $(\beta, Q_e)$ parameter space using the EHT shadow observations of M87*\cite{M87_1_2019,M87_2_2019,M87_3_2019,M87_4_2019,M87_5_2019,M87_6_2019}. The EHT collaboration measured the shadow diameter of M87* as $D_{sh} = 11.0 \pm 1.5$ [M], which provides constrained ranges for the shadow radius $R_{sh} = D_{sh}/2$: $R_{sh} \in [4.75, 6.25]$ ($1\sigma$) and $R_{sh} \in [4, 7]$ ($2\sigma$)\cite{Jafarzade:2025nbe}. We present the parameter space $(Q_e, \beta)$ in Fig. \ref{phase2} corresponding to these constraints, and it is noted that the order of magnitude is constrained to $\beta \sim 0.1$.
\subsection{dyonic charged case}
From the Eq. (\ref{eff}) we can obtain the effective metric of dyonic charged black hole, which reads
\begin{equation}
	\begin{split}		
		w_\text{dy}(r)&=A_{\text{dy}}(r),\\
		h_\text{dy}(r)&=\frac{8 \pi  r^{12}}{225 a \eta ^2 Q_e^2 \epsilon^2-180 a \eta  Q_e^2 r^4 \epsilon+4 r^8 \left(9 aQ_e^2-8 a Q_m^2+8 r^4\right)},\\
		u_\text{dy}(r)&=\frac{r^{10} \left(17 a Q_m^2-8 r^4\right)}{8a \left(50 \eta ^2 Q_e^2 \epsilon^2+r^8 \left(8 Q_e^2-9 Q_m^2\right)-40 \eta  Q_e^2 r^4 \epsilon\right)+64 r^{12}}+\frac{9r^2}{8}.
	\end{split}
\end{equation}
\par For the dyonic EH electromagnetic field in Table. \ref{t2}, similar to the magnetic charge, increasing electric and magnetic charges significantly reduces \( r_h \), \( r_{ph} \), and \( r_{\text{ISCO}} \). As \( \beta \) increases, \( r_h \) first decreases then increases, while the photon sphere and ISCO approach the event horizon. 

\begin{table}
	\setlength{\tabcolsep}{3mm}
	\begin{center}		
		\begin{tabular}[b]{c|cccccc}
			\midrule[2pt]
			$Q_e$ & 0 & 0.2 & 0.4 & 0.6 & 0.8 & 0.9  \\
			\toprule[1pt]
			$r_{\text{ph}}$ & 2.82288 & 2.79229 & 2.69583 & 2.51489 &2.18557& 1.86055  \\
			$r_{\text{ISCO}}$&5.60664 &5.53983 & 5.33155&4.95207 &4.31323 &3.80142\\
			\midrule[1pt]
		\end{tabular}
	\end{center}
	\centering
	\caption{Data of $r_{\text{h}}$, $r_{\text{ph}}$, $b_{\text{ph}}$ of black holes in $f(R,T)$ gravity coupled with EH dyonic electromagnetic field, for different $Q_e$  with fixed $M=1,Q_m=0.7,  \beta=0.6,\eta=2\alpha/225\pi ,a=0.001786$. }
	\label{t2}
\end{table}
\begin{figure}
	\begin{center}
		\subfigure[$\beta=-0.1,Q_e=0.7$]{\includegraphics[width=4cm]{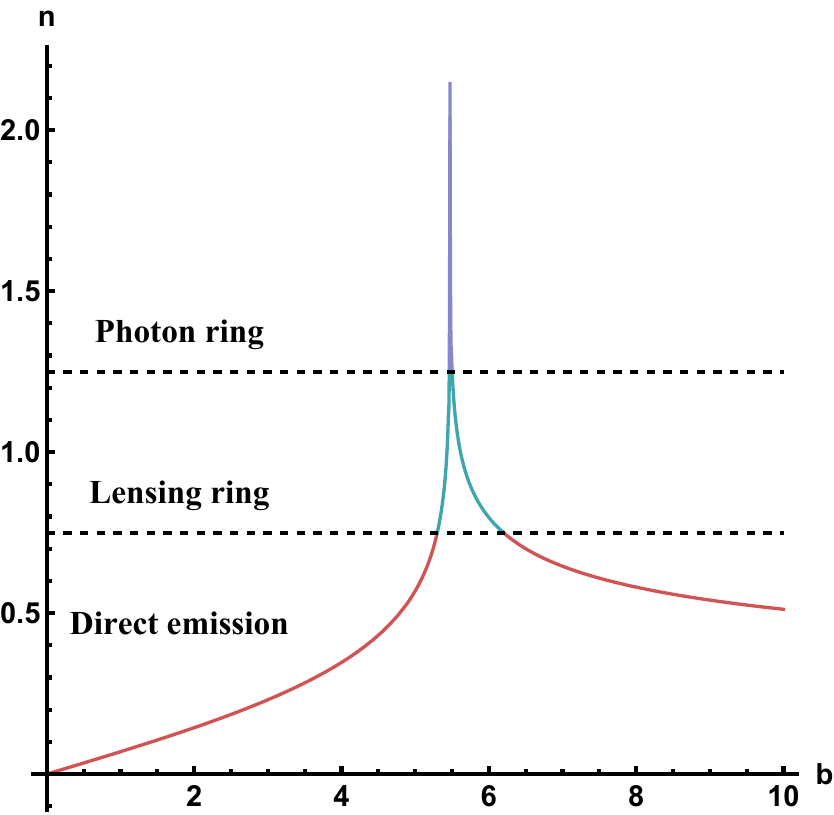}\label{gph1}}
		\subfigure[$\beta=-0.2,Q_e=0.7$]{\includegraphics[width=4cm]{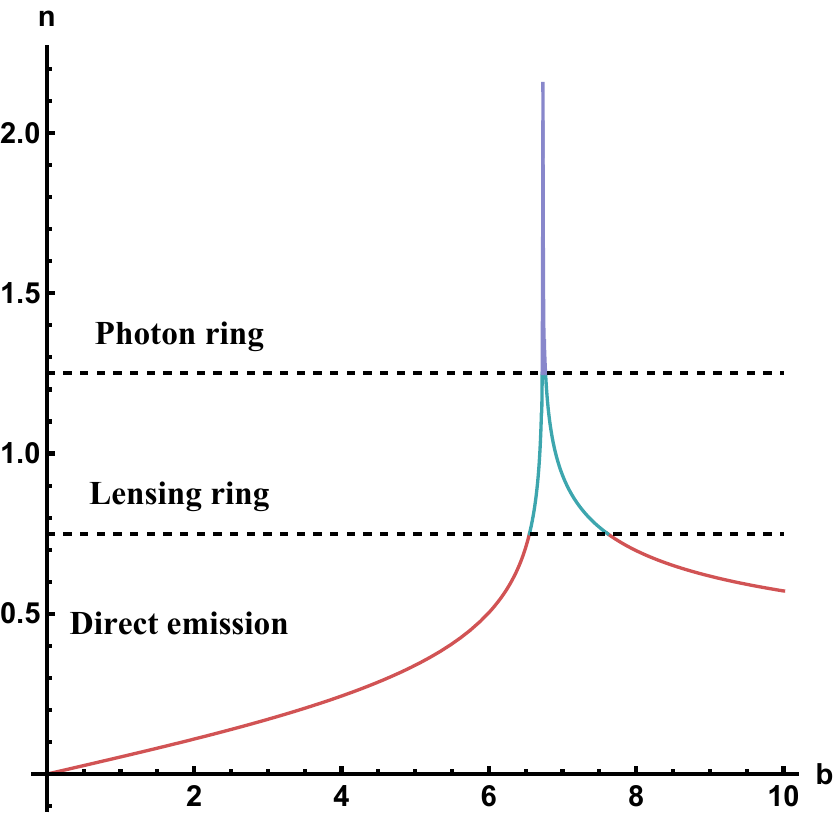}\label{gph11}}
		\subfigure[$Q_e=0.2,Q_m=0.2$]{\includegraphics[width=4cm]{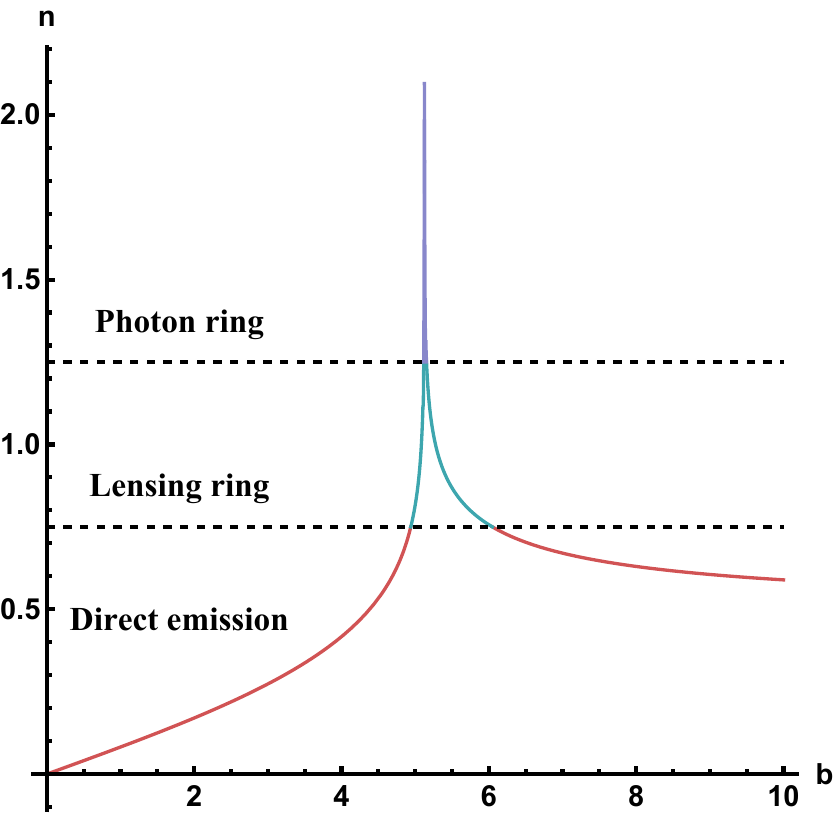}\label{gph2}}
		\subfigure[$Q_e=0.7,Q_m=0.7$]{\includegraphics[width=4cm]{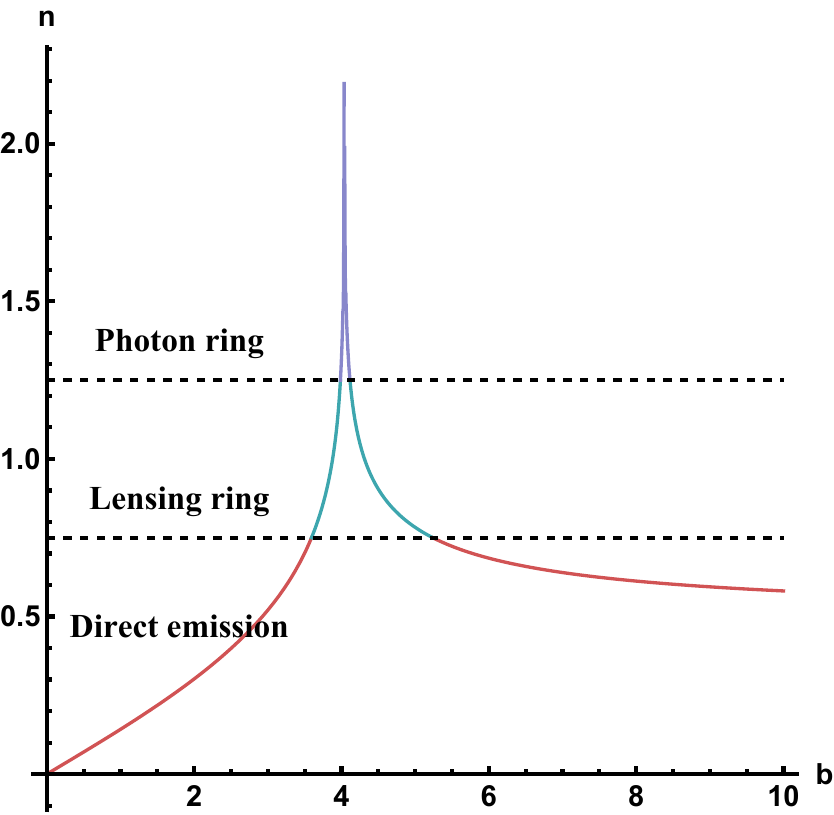}\label{gph3}}
		\subfigure[$\beta=-0.1,Q_e=0.7$]{\includegraphics[width=4cm]{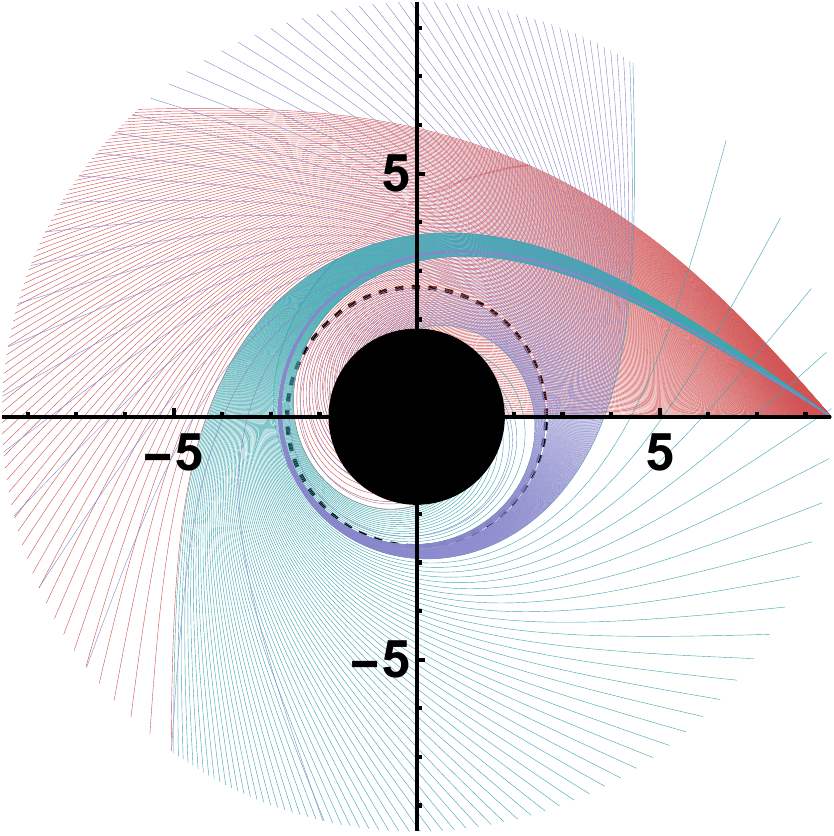}\label{g1}}	
		\subfigure[$\beta=-0.2,Q_e=0.7$]{\includegraphics[width=4cm]{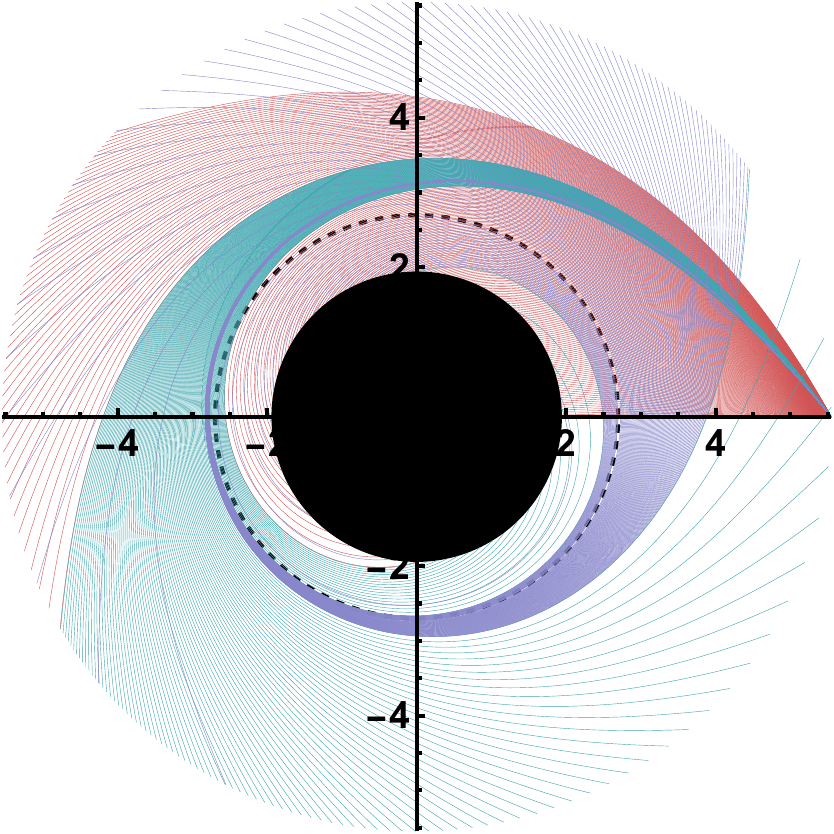}\label{g11}}	
		\subfigure[$Q_e=0.2,Q_m=0.2$]{\includegraphics[width=4cm]{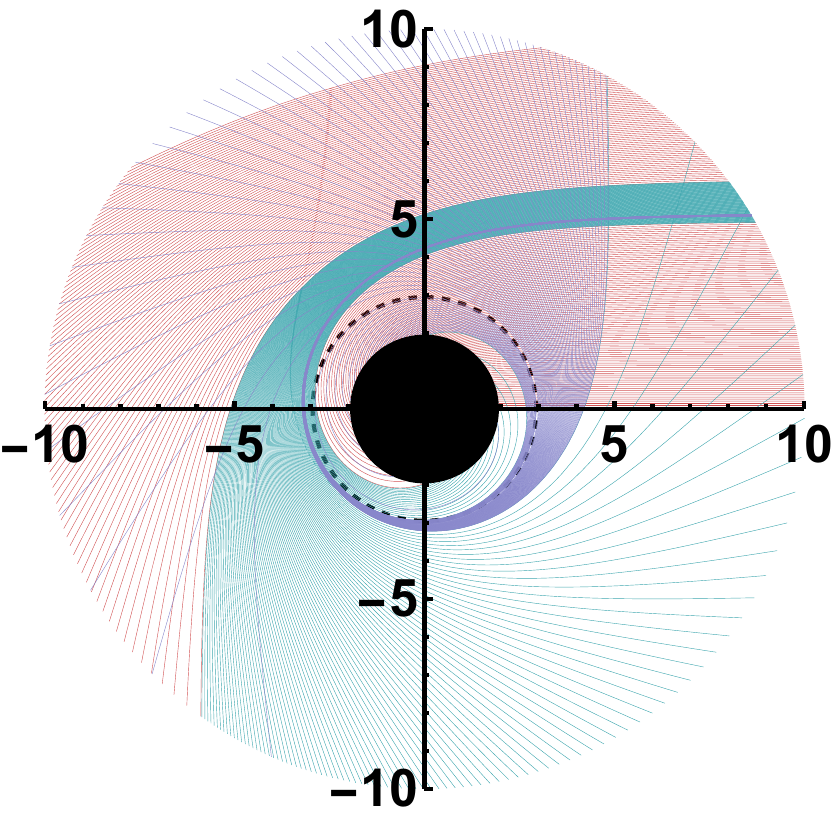}\label{g2}}
		\subfigure[$Q_m=0.7,Q_m=0.7$]{\includegraphics[width=4cm]{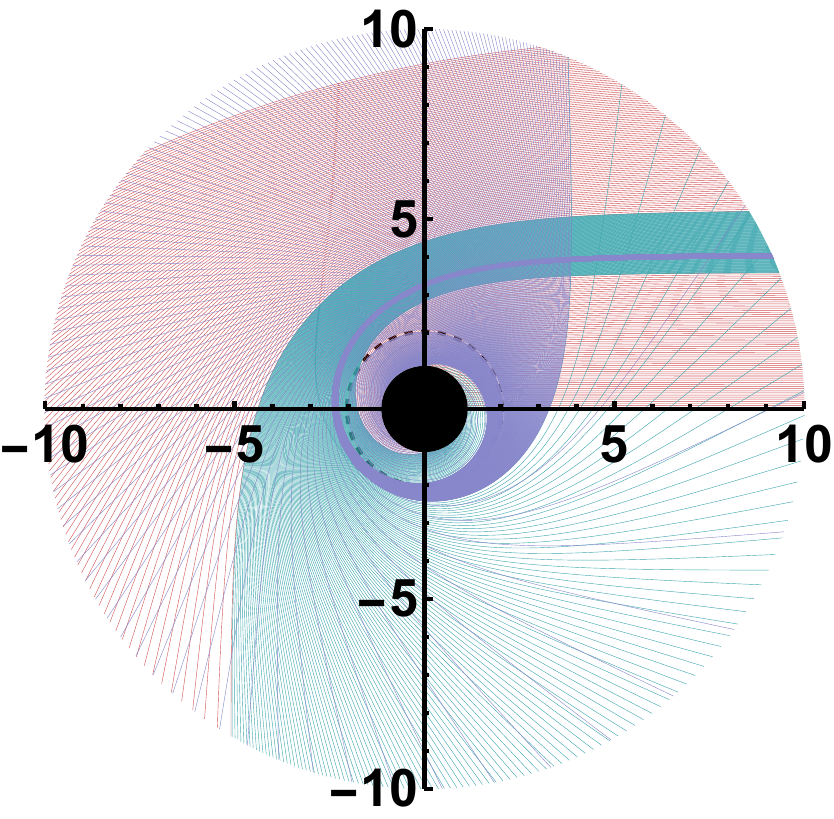}\label{g3}}
	\end{center}
	\caption{Geodesics structure of electrically and dyonic charged black hole in $f(R,T)$ gravity coupled with EH field. Upper line are the number of orbits $n$ with respect to impact parameter $b$. Lower line are the geodesics figure. All fixed $a=2$ and $M=1$, and we fixed $\beta=0.6$ for dyonic case in the right columns.}
	\label{geo2}
\end{figure}
	\par The geodesics figure are shown in the right two columns of Fig. \ref{geo2}, increasing magnetic and electric charges causes photons with lower impact parameter $b$ to achieve a higher number of orbital loops $n$, with a significant increase in the proportion of photons in the photon ring and lensing ring. The Increasing coupling constant \( \beta \) significantly enhances the proportion of the photon ring and lensing ring, and photon trajectories with smaller impact parameters exhibit highly pronounced light deflection. This indicates that the introduction of electric charge amplifies the matter field's influence on the gravitational field, particularly evident at higher coupling strengths.
\section{Conclusion and discussion\label{s8}}
\par In this study, we investigated black hole solutions in $f(R,T)$ gravity coupled with Euler-Heisenberg nonlinear electrodynamics, utilizing both Born-Infeld and QED interpretations to derive analytical metrics for magnetic, electric, and dyonic charges. A distinct feature of the electric solution is the emergence of an effective cosmological constant determined by the coupling parameter $\beta$, which dictates an AdS or dS spacetime structure, unlike the asymptotically flat magnetic and dyonic cases. Our analysis of horizons and photon dynamics demonstrates that increasing charge strengths and the coupling magnitude generally reduces the shadow radius across all solutions. Furthermore, the interplay between modified gravity and nonlinear electrodynamics significantly impacts the horizon structure—such as the formation of cosmological horizons in dS spacetimes—and enhances the observational signatures of photon and lensing rings in the strong-field regime.
\par Our work differs from previous studies in Einstein–Euler–Heisenberg (EEH) theory \cite{Magos:2023nnb,Ruffini:2013hia,Meng:2021cgb,Amaro:2020xro,Amaro:2022yew,Amaro:2022del,Breton:2023bwf,Amaro:2023ull,Cotton:2021tfl,Yajima:2000kw,Vagnozzi:2022moj,Li:2021ygi,Magos:2020ykt,Blazquez-Salcedo:2020caw,Karakasis:2022xzm,Maceda:2018zim,Gutierrez-Cano:2024oon,Rehman:2023hro,Breton:2021mju,Allahyari:2019jqz} by introducing a non-minimal matter–geometry coupling through $f(R, T)$ gravity, where the trace $T$ of the energy-momentum tensor directly influences spacetime evolution. This  leads to the following new features: 
	\newline(i) modified black hole metric functions, such as a $\beta$-dependent correction in the $1/r^6$ term for magnetically charged solutions (Section \ref{s3}) and an $r^2$ term acting as an effective cosmological constant in electrically charged cases (Section \ref{s7}), resulting in AdS- or dS-like spacetimes; 
	\newline(ii) altered horizon structures, including the emergence of cosmological horizons for negative $\beta$ and changes in extremal black hole charges; 
	\newline(iii) breaking of electromagnetic duality in the dyonic case due to $f(R, T)$ effects, incorporating higher-order vacuum polarization corrections; 
	\newline(iv) qualitative analysis of energy conditions under this coupling (Section \ref{s5}); 
	\newline(v) detailed investigation of effective metrics, photon trajectories, and innermost stable circular orbits (ISCO), with visual representations of geodesics for varying coupling parameters $\beta$ and charges $Q_e$, $Q_m$ (Section \ref{s6}). These additions provide insights into how matter-geometry interactions in modified gravity could manifest in strong-field environments, potentially observable through black hole shadows or gravitational wave polarizations.
\begin{acknowledgments}
\par The authors are grateful to Wei Hong, Aoyun He, Yadong Xue, Guohe Li and Mian Zhu for useful discussions and insightful
suggestions.  This work is supported by the National Natural Science Foundation of China (NSFC) with Grants No. 12175212. And it is finished on the server from Kun-Lun in College of Physics, Sichuan University.
\end{acknowledgments}
\appendix
\section{Kretschmann scalar of dyonic EH black hole in $f(R,T)$ gravity\label{ap}}
\par Here we give the Kretschmann scalar of dyonic EH black hole in $f(R,T)$ gravity, reads
\begin{equation}
	\begin{split}
		&\mathcal{K}_{dyonic}=\frac{1399609375 a^2 \left(2 Q_e^2+5 Q_m^2\right)^8 (\beta -2 \pi )^2 \eta ^8 Q_e^8}{3035648 \pi ^2 r^{48}}+\frac{239 a^2 \beta ^2 Q_e^8}{450 \pi ^2 r^{16}}+\frac{478 a^2 Q_e^8}{225 r^{16}}-\frac{478 a^2 \beta  Q_e^8}{225 \pi  r^{16}}\\&-\frac{852734375 a^2 \left(2 Q_e^2+5 Q_m^2\right)^7 (\beta -2 \pi )^2 \eta ^7 Q_e^8}{565488 \pi ^2 r^{44}}+15625 a^2 \left(2 Q_e^2+5 Q_m^2\right)^6 (2 \pi -\beta ) \times\\&\frac{\left(-1457769314 \beta  Q_e^2+2915538628 \pi  Q_e^2+1887280398 \pi  Q_m^2-733942377 Q_m^2 \beta \right) \eta ^6 Q_e^6}{10497719232 \pi ^2 r^{40}}\\&+\frac{1673 a^2 Q_m^2 \beta ^2 Q_e^6}{225 \pi ^2 r^{16}}+\frac{956 a^2 Q_m^2 Q_e^6}{25 r^{16}}-\frac{7648 a^2 Q_m^2 \beta  Q_e^6}{225 \pi  r^{16}}\\&-\frac{625 a^2 \left(2 Q_e^2+5 Q_m^2\right)^5 (2 \pi -\beta ) \left(-1762498 \beta Q_e^2+3524996 \pi  Q_e^2+6849522 \pi  Q_m^2-2663703 Q_m^2 \beta \right) \eta ^5 Q_e^6}{612612 \pi ^2 r^{36}}\\&+\frac{4541 a^2 Q_m^4 \beta ^2 Q_e^4}{450 \pi ^2 r^{16}}+\frac{478 a^2 Q_m^4 Q_e^4}{3 r^{16}}-\frac{478 a^2 Q_m^4 \beta Q_e^4}{5 \pi  r^{16}}-\frac{25 a^2 \left(2 Q_e^2+5 Q_m^2\right)^3\eta ^3 Q_e^4}{83538 \pi ^2 r^{28}} \times\\&(\left(1073498 Q_e^4+5402474 Q_m^2 Q_e^2+1170771 Q_m^4\right) \beta ^2-4 \pi  \left(1073498 Q_e^4+6174256 Q_m^2Q_e^2+3236877 Q_m^4\right) \beta\\& +4 \pi ^2 \left(1073498 Q_e^4+6946038 Q_m^2 Q_e^2+5604699 Q_m^4\right))+\frac{24856 a^2 Q_m^6 \beta Q_e^2}{75 \pi  r^{16}}-\frac{2868 a^2 Q_m^6 Q_e^2}{25 r^{16}}\\&-\frac{1673 a^2 Q_m^6 \beta ^2 Q_e^2}{15 \pi ^2 r^{16}}-608 a^2 \left(2 Q_e^2+5 Q_m^2\right) \left(-2 \beta  Q_e^2+4 \pi  Q_e^2+18 \pi Q_m^2-7Q_m^2 \beta \right)\times\\&\frac{ \left(2 \pi  \left(Q_e^4+9 Q_m^2 Q_e^2-3 Q_m^4\right)-\left(Q_e^4+7 Q_m^2Q_e^2-15Q_m^4\right) \beta \right) \eta Q_e^2}{135 \pi ^2 r^{20}}+\frac{239 a^2 Q_m^8 \beta ^2}{2 \pi ^2 r^{16}}+\frac{a^2 Q_e^2 \left(2 Q_e^2+5 Q_m^2\right)^2\eta ^2}{9828 \pi ^2 r^{24}}\times\\&(\left(688438Q_e^6+5179741 Q_m^2 Q_e^4+2533753 Q_m^4 Q_e^2-5410125 Q_m^6\right) \beta ^2-4 \pi  (688438Q_e^6+5919704Q_m^2 Q_e^4\\&+6437511 Q_m^4 Q_e^2-4018950Q_m^6) \beta +4 \pi ^2 \left(688438Q_e^6+6659667 Q_m^2 Q_e^4+10926657 Q_m^4 Q_e^2-1391175 Q_m^6\right)) \\&+\frac{125 a^2 Q_e^4 \left(2 Q_e^2+5 Q_m^2\right)^4}{4013833824 \pi ^2 r^{32}} (\left(30255060830 Q_e^4+91494873274 Q_m^2 Q_e^2+6036696855Q_m^4\right) \beta ^2\\&-4 \pi  \left(30255060830 Q_e^4+104565569456 Q_m^2 Q_e^2+17900307369Q_m^4\right) \beta \\&+4 \pi ^2 \left(30255060830 Q_e^4+117636265638 Q_m^2 Q_e^2+31463714223 Q_m^4\right)) \eta ^4+\frac{432}{54 r^{24}} \times\\&((7 \left(Q_e^2+Q_m^2\right)^2-12 r \left(Q_e^2+Q_m^2\right)+6 r^2) r^{16}-576 Q_e^2 \left(2 Q_e^2+5 Q_m^2\right) \left(19 \left(Q_e^2+Q_m^2\right)-14 r\right) \eta  r^{12}\\&+24 Q_e^2 \left(2 Q_e^2+5 Q_m^2\right)^2 \left(943 Q_e^2+465 Q_m^2-330 r\right) \eta ^2 r^8-24320 Q_e^4 \left(2Q_e^2+5Q_m^2\right)^3 \eta ^3 r^4\\&+13025 Q_e^4 \left(2 Q_e^2+5 Q_m^2\right)^4 \eta ^4)+\frac{478 a^2 Q_m^8}{25 r^{16}}-\frac{478 a^2Q_m^8 \beta }{5 \pi  r^{16}}+\frac{a}{27567540 \pi  r^{36}}\times\\&((14702688 \left(19 \left(Q_e^2+Q_m^2\right)-14 r\right) \left(2 \pi  \left(Q_e^4+9 Q_m^2 Q_e^2-3 Q_m^4\right)-\left(Q_e^4+7 Q_m^2Q_e^2-15Q_m^4\right) \beta \right) r^{24}\\&-2450448 Q_e^2 \left(2Q_e^2+5 Q_m^2\right) (2 \pi  \left(1169 Q_e^4+7266 Q_m^2 Q_e^2+3468 Q_m^4-330 \left(2 Q_e^2+9 Q_m^2\right) r\right)\\&+\left(-1169 Q_e^4+\left(660 r-5858Q_m^2\right) Q_e^2+330 \left(Q_m^4+7 r Q_m^2\right)\right) \beta ) \eta  r^{20}\\&+44880 Q_e^2 \left(2 Q_e^2+5 Q_m^2\right)^2 (6 \pi  \left(99546Q_e^4+352605 Q_m^2 Q_e^2+73243Q_m^4-20250 \left(2 Q_e^2+3 Q_m^2\right) r\right)\\&+\left(-298638 Q_e^4-861445 Q_m^2 Q_e^2+4305 Q_m^4+20250 \left(6 Q_e^2+7 Q_m^2\right) r\right) \beta ) \eta ^2 r^{16}\\&-6600 Q_e^4 \left(2 Q_e^2+5 Q_m^2\right)^3 (2 \pi  \left(5310637 Q_e^2+9445479 Q_m^2-1296750 r\right)+\\&\left(-5310637 Q_e^2-7763567 Q_m^2+1296750 r\right) \beta ) \eta ^3 r^{12}+3750 Q_e^4 \left(2Q_e^2+5 Q_m^2\right)^4 \times \\&(-14384659 \beta  Q_e^2+28769318 \pi  Q_e^2+1836 \pi  \left(10312 Q_m^2-1495 r\right)+306 \left(4485 r-25513 Q_m^2\right) \beta ) \eta ^4 r^8\\&-46911353125 Q_e^6 \left(2 Q_e^2+5 Q_m^2\right)^5 (2 \pi -\beta ) \eta ^5 r^4+18305015625 Q_e^6 \left(2 Q_e^2+5Q_m^2\right)^6 (2 \pi -\beta ) \eta ^6)).
	\end{split}
\end{equation}
\normalem
\nocite{*}
\bibliography{ref21}
\end{document}